\documentclass[10pt,journal,compsoc]{IEEEtran}

\usepackage{amsmath,amsfonts}
\usepackage{algorithmic}
\usepackage{algorithm}
\usepackage{array}
\usepackage[caption=false,font=normalsize,labelfont=sf,textfont=sf]{subfig}
\usepackage{textcomp}
\usepackage{stfloats}
\usepackage{url}
\usepackage{verbatim}
\usepackage{graphicx}
\usepackage{cite}
\usepackage{booktabs} 
\usepackage{tabularx} 
\usepackage{threeparttable}
\usepackage{color}
\usepackage[normalem]{ulem}
\usepackage{multirow}
\usepackage[numbers,sort&compress]{natbib}

\begin{document}
%
\title{DFG-PCN: Point Cloud Completion with Degree-Flexible Point Graph }
%
%
%
%
\author{Zhenyu~Shu,
Jian~Yao*,
Shiqing~Xin
\IEEEcompsocitemizethanks{
	\IEEEcompsocthanksitem Zhenyu Shu is with School of Computer Science and Technology, Zhejiang Sci-Tech University, Hangzhou, PR China. He is also with School of Computer and Data Engineering, NingboTech University, Ningbo, PR China.
	\protect\\
	E-mail: shuzhenyu@nit.zju.edu.cn (Zhenyu Shu)
	\IEEEcompsocthanksitem Jian Yao is with School of Computer Science and Technology, Zhejiang Sci-Tech University, Hangzhou, PR China. Corresponding author.
	\protect\\
	E-mail: jianyao\_paper@163.com (Jian Yao)
	\IEEEcompsocthanksitem Shiqing Xin is with School of Computer Science and Technology, ShanDong University, Jinan, PR China.
	}
\thanks{Manuscript received month day, year; revised month day, year.}
}

%
%

\markboth{IEEE transactions on visualization and computer graphics,~Vol.~XX, No.~X, Month~Year}%
{Shell \MakeLowercase{\textit{et al.}}: Bare Demo of IEEEtran.cls for Computer Society Journals}
%



\IEEEtitleabstractindextext{%
\begin{abstract}
	Point cloud completion is a vital task focused on reconstructing complete point clouds and addressing the incompleteness caused by occlusion and limited sensor resolution. Traditional methods relying on fixed local region partitioning, such as $k$-nearest neighbors, which fail to account for the highly uneven distribution of geometric complexity across different regions of a shape.  This limitation leads to inefficient representation and suboptimal reconstruction, especially in areas with fine-grained details or structural discontinuities. This paper proposes a point cloud completion framework called Degree-Flexible Point Graph Completion Network (DFG-PCN). It adaptively assigns node degrees using a detail-aware metric that combines feature variation and curvature, focusing on structurally important regions. We further introduce a geometry-aware graph integration module that uses Manhattan distance for edge aggregation and detail-guided fusion of local and global features to enhance representation. Extensive experiments on multiple benchmark datasets demonstrate that our method consistently outperforms state-of-the-art approaches.
\end{abstract}

}

\maketitle

\IEEEdisplaynontitleabstractindextext

%
\IEEEpeerreviewmaketitle

\IEEEraisesectionheading{\section{Introduction}}
\IEEEPARstart{P}{oint} clouds~\cite{cadena2016past,reddy2018carfusion,rusu2008towards} serve as a widely adopted and easily accessible data format for representing 3D objects, playing a crucial role in advancing research within computer vision, particularly in the understanding of 3D scenes and object structures~\cite{10471330, han2020shapecaptioner, ShuYu-9572, ShuShen-9376}. The point clouds obtained from real-world scenarios are frequently characterized by significant sparsity and incompleteness. These challenges arise primarily due to constraints such as restricted viewpoints, occlusion caused by object self-geometry, and the limited resolution of sensing equipment. Therefore, recovering complete point clouds is an essential downstream task, primarily aimed at preserving the observed details, inferring missing parts, and densifying sparse surfaces~\cite{liang2018deep,le2018pointgrid}.

In recent years, deep learning-based methods have been developed for point cloud completion. Notably, with the success of PointNet~\cite{qi2017pointnet} and PointNet++~\cite{qi2017pointnet++} in point cloud deep learning, most methods~\cite{yuan2018pcn,xiang2021snowflakenet,zhou2022seedformer,chen2023anchorformer} directly generate complete point clouds based on 3D coordinates. Due to point cloud data's unordered and unstructured nature, learning fine-grained geometric and structural features is essential for generating plausible shapes.

\begin{figure}[t]
	\centering
	\includegraphics[width=1\columnwidth]{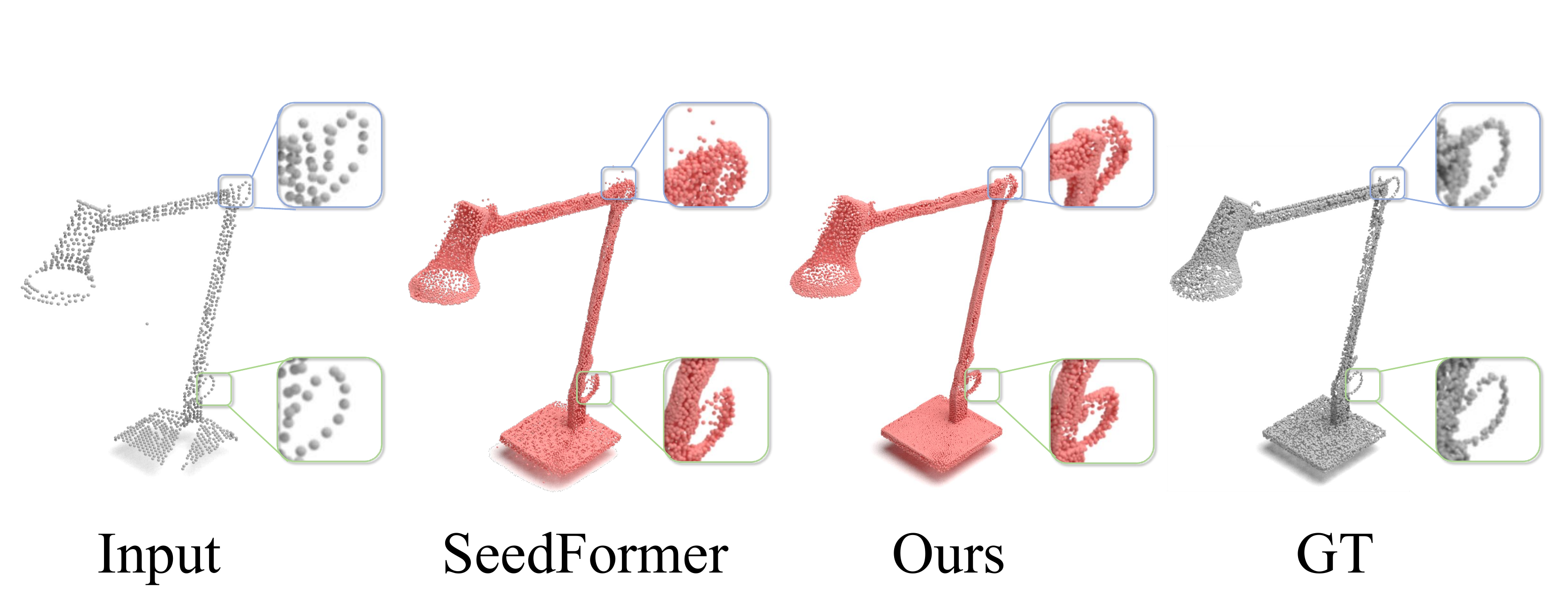}
	\caption{A comparative analysis of point cloud completion techniques is presented. The input comprises 2048 points, while the ground truth contains 16,384 points. Unlike methods like SeedFormer, our approach reconstructs complete shapes with 16,384 points, offering superior geometric details, including clearly defined smooth surfaces (marked in green) and reduced, well-distributed noise (marked in blue).}
	\label{fig:1}
\end{figure}

Recent works~\cite{10232862,wang2024pointattn,yan2022fbnet,xie2020grnet,chen2023anchorformer,yu2021pointr,wen2020point} on point cloud completion are typically formulated as a generation problem, primarily utilizing an encoder-decoder structure to achieve this goal. In this framework, the input partial point cloud is encoded into a global feature vector and then decoded to recover a complete point cloud from low to high resolution. However, many existing methods face challenges due to the inherent discreteness of point clouds and the unstructured nature of local point predictions, making preserving structural consistency within local patches challenging. Accurately capturing fine-grained geometric details and structural characteristics in complete shapes remains a significant challenge, especially in regions with smooth surfaces, sharp edges, and corners, as illustrated in Figure~\ref{fig:1}.

Additionally, point cloud completion tasks exhibit a clear imbalance in reconstruction requirements: while only a small number of high-frequency points require intricate reconstruction, the majority of points lie in flat, low-frequency regions that largely retain their geometric features. To address this imbalance, an ideal point cloud completion method should focus more on detail-rich areas and pay less attention to flat, feature-sparse regions of the point cloud. However, existing methods~\cite{zhou2022seedformer,yu2021pointr,10232862,qi2017pointnet++} based on $k$-nearest neighbors (KNN) treat all nodes equally. In other words, all nodes have the same predefined degree $k$ without considering the imbalanced nature of point cloud completion.

When we analyze the classical operation paradigms in point cloud completion from a graph perspective, we find that the degree-equivalent property also exists in convolution and attention mechanisms. Each point aggregates a fixed number of neighboring points in these paradigms, regardless of the point cloud's content, which manifests as ``equal degrees" in graph terminology. In point cloud completion, the fixed equal degree of nodes or points mismatches their unequal reconstruction demands, thereby impacting the overall performance of point cloud completion.

To resolve this problem, we present the Degree-Flexible Point Graph (DFG) model designed for point cloud completion, allowing for adaptive adjustment of node degrees. This methodology represents fine-grained local and overarching global geometric features of point cloud data, thereby boosting completion performance. Given a partial point cloud, the DFG framework uses a feature extraction mechanism to reduce the point cloud's scale and obtain initial point features. These extracted features are inputs to a seed generator to produce a sparse point cloud representing the complete shape. Subsequently, an encoder with a flexible point cloud graph structure takes the generated sparse point cloud as input. It iteratively enhances its resolution until the target resolution is achieved, thereby further optimizing the point cloud completion.

To validate the proposed method's effectiveness, we evaluated four benchmark datasets: PCN~\cite{yuan2018pcn}, ShapeNet-55, ShapeNet-34~\cite{yu2021pointr}, and KITTI~\cite{geiger2013vision}. Experimental results demonstrate that the proposed approach can recover detailed and plausible shapes for the missing parts, achieving satisfactory outcomes. Our contributions are threefold:
\begin{itemize}
	\item We introduce a new DFG-PCN model for point cloud completion, designed to effectively learn both local and global features while generating detailed point cloud predictions.
	\item We propose a detail-aware degree allocation strategy that leverages feature variation and curvature to assign more connections to structurally important regions adaptively.
	\item We develop a geometry-aware graph integration module combining Manhattan-distance-based aggregation and detail-guided local-global feature fusion, improving structural consistency and detail preservation.
	
\end{itemize}

The organization of this paper is as follows: Section \ref{rw} reviews prior work in the area of 3D point cloud completion. Section \ref{md} details the flexible degree graph network framework proposed in this study. In Section \ref{exp}, we provide comprehensive experimental results and benchmark comparisons. Finally, Section \ref{lf} outlines the method's limitations and explores potential future research avenues, and Section \ref{cl} offers a summary of conclusions.

\begin{figure*}[t]
	\centering
	
	\includegraphics[width=2\columnwidth]{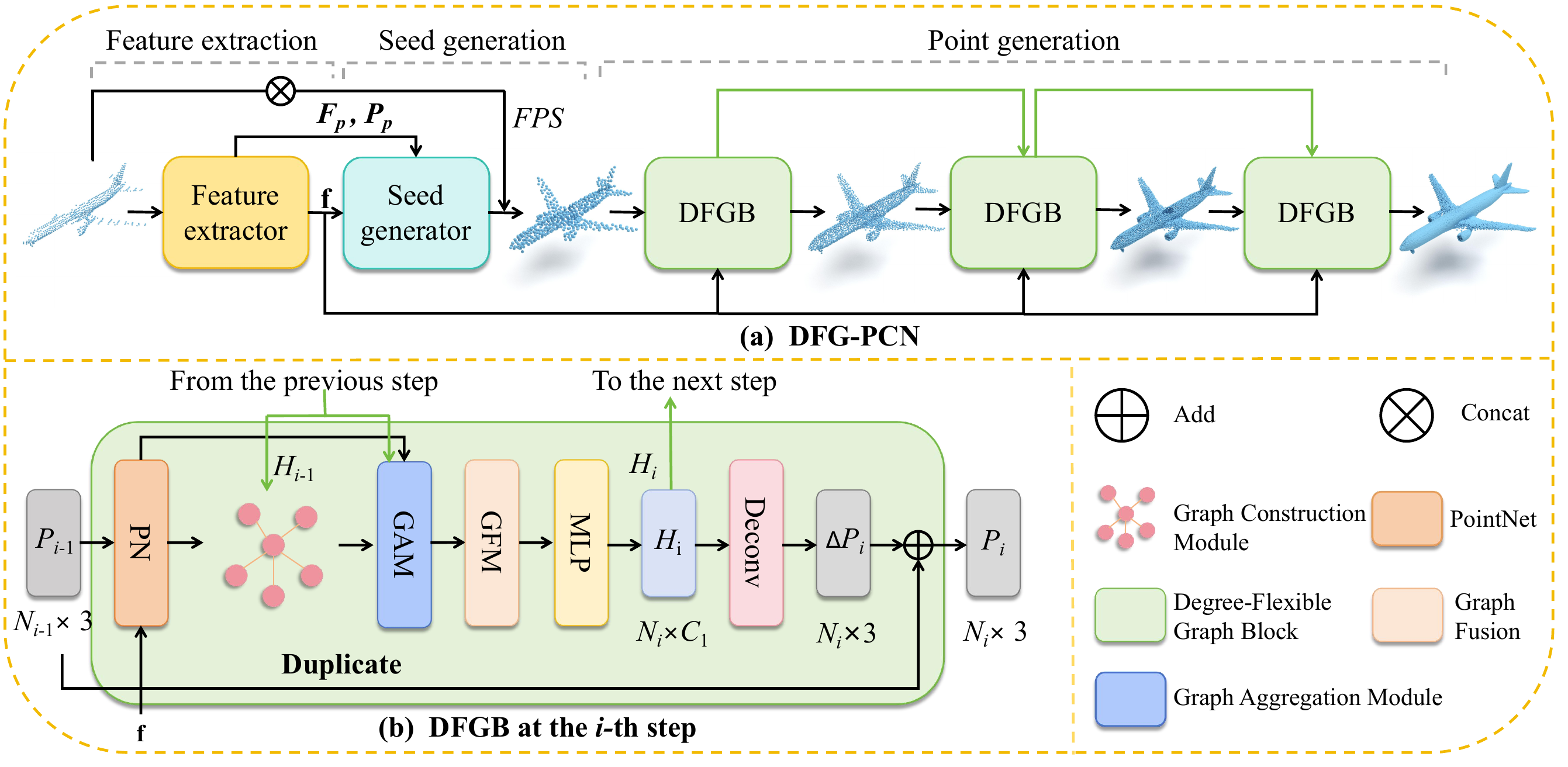}
	\caption{(a) The overall architecture of DFG-PCN consists of three blocks: feature extractor, seed generation, and point generation. (b) The details of the Degree-Flexible Graph Block are as follows: PointNet, Graph Construction Module, Graph Aggregation Module, Graph Fusion Module, Deconvolution and MLP. Note that $N_{i}$ is the number of points,  and $C_{1}$ is the number of point feature channels.}
	\label{fig:2}
\end{figure*}

\section{Related Work}\label{rw}
Conventional point cloud completion approaches~\cite{berger2014state,hu2019local,nguyen2016field} depend on manually designed features, including properties like surface smoothness or symmetry axes, to achieve the reconstruction of complete point cloud shapes. Other approaches~\cite{sung2015data,shen2012structure} extract local geometric information from a large set of basic 3D structured shapes to build structural local priors for partial shapes, aiding in completing complex shapes. These methods depend on existing structured data distributions, making it challenging to cover all possible cases comprehensively.

Thanks to the rapid progress in deep learning, modern techniques now leverage deep models for point cloud completion. The literature in this domain can be divided into two primary types: voxel-based methods and point cloud-based approaches for shape completion.

\subsection{Voxel-based Shape Completion}
Recent advances in deep neural networks have enabled effective encoding of point cloud geometry for shape completion. Inspired by the success of CNNs in 2D image analysis, some works extend this to 3D using voxel-based representations and 3D CNNs~\cite{maturana2015voxnet, wang2017cnn,wu2018learning,dai2017shape,han2017high,le2018pointgrid}.

For example, Han et al. ~\cite{han2017high} and GRNet~\cite{xie2020grnet} apply 3D CNNs to learn shape structures and complete point clouds via voxel grids. However, these methods face high computational costs and struggle to capture fine geometric details due to information loss during voxelization and the limitations of fixed grid structures.

\subsection{Point-based Shape Completion}
With the success of PointNet~\cite{qi2017pointnet} indirectly processing 3D coordinates, point-based methods have gradually become the mainstream solution for point cloud completion tasks, achieving significant progress. Building on previous work, several point-based methods have been proposed. Some of these approaches transform local regions into graph structures, subsequently processed by graph convolution layers~\cite{wang2019dynamic,zhao2019pointweb,simonovsky2017dynamic}. Point feature learning has been inspired by the success of image~\cite{dosovitskiy2020image} and video~\cite{liu2022video} transformers, leading to the adoption of self-attention mechanisms~\cite{wang2024pointattn,yu2021pointr}. Additionally, adversarial learning has been used to improve the realism of generated shapes, while cross-modal feature learning facilitates shape reconstruction from images~\cite{zhu2023svdformer}. 

TopNet~\cite{tchapmi2019topnet} presents a one-stage architecture that models point cloud generation as a tree-like process, where a parent feature is split to generate multiple child features. Although this method improves the decoder by incorporating a rooted tree structure, which enhances topological reconstruction, the generated point features still lack precise shape information in the incomplete regions, limiting their ability to impose strong constraints.

Two-stage point cloud completion frameworks can improve performance by applying more constraints during the progressive generation from coarse to fine, thereby enhancing the accuracy of the resulting point clouds. PCN~\cite{yuan2018pcn} was the first learning-based approach to use an encoder-decoder design. This method recovers point clouds through two stages: first, the encoder extracts global features from the input data, and then a MultiLayer Perceptron (MLP) generates an initial coarse prediction, which is subsequently refined into a more detailed result. Approaches such as CDN~\cite{wang2020cascaded} have expanded this method by increasing the number of generation stages, leading to superior performance. Several other methods~\cite{xiang2021snowflakenet, zhou2022seedformer, wang2024pointattn} have achieved significant improvements by introducing additional stages, enhancing feature extraction, or optimizing the generation process structure. PointAttN~\cite{wang2024pointattn} replaces the conventional KNN-based local feature extraction with an attention mechanism, enabling more effective modeling of global dependencies within point cloud structures. PPCL~\cite{fei2025point} introduces a multi-layer contrastive learning framework aligning encoder-decoder features to optimize geometric consistency, enhancing point cloud completion accuracy via spatial-channel Transformer upsampling. By introducing a latent diffusion-based framework~\cite{rombach2022high} with dual-pathway VAE modeling and cross-attention-driven global-local fusion, PointLDM~\cite{chen2024learning} provides new insights into point cloud completion. Recent methods~\cite{zhou2022seedformer,wang2024pointattn,fei2025point} incorporate global-local fusion via attention or hierarchical encoding. However, most apply uniform fusion strategies, failing to distinguish the relative importance of different spatial regions, particularly in detail-rich areas.

Our approach is part of the field of shape completion methods based on point clouds. Our DFG approach utilizes flexible graphs to adaptively focus on regions with richer details, thereby enhancing the precision of point cloud completion.

\section{Methodology}\label{md} 
The DFG framework is depicted in Figure~\ref{fig:2}. Given an incomplete and sparse point cloud $ P\in \mathbb{R} ^{N_{p} \times 3}$ as input, the aim is to recover the missing parts and generate a full, dense point cloud $P_{3} \in \mathbb{R} ^{N_{3} \times 3}$. This system follows a widely used two-stage point cloud completion methodology. It comprises three main components: a feature extraction module, a seed generation module, and a degree-flexible point graph module.

\subsection{DFG-PCN}
Figure~\ref{fig:2} depicts the overall structure of DFG-PCN, which includes a feature extractor, a seed generator, and a point generation module.

\textbf{Feature extractor.} The feature extractor is responsible for extracting and integrating the input shape vector with the features of individual points in the cloud. Denote the input point cloud as $P\in \mathbb{R}^{N \times 3}$, where $N$ represents the number of points, and each point corresponds to a 3D coordinate. Using hierarchical aggregation, downsampling, and max-pooling operations, we generate the shape vector \textbf{f} $\in \mathbb{R}^{N}$, as well as the downsampled partial points $P_{p}\in \mathbb{R}^{N_{p}\times 3}$ and their associated feature vectors $F_{p}\in \mathbb{R}^{N_{p}\times C}$, where $N_{p}$ refers to the number of points in the downsampled partial point cloud. We apply three layers of set abstraction from~\cite{qi2017pointnet++}, which aggregate point features from local to global levels. In addition, the Point Transformer~\cite{zhao2021point} is integrated to improve the capturing and incorporation of local shape context.

\textbf{Seed generator.} The objective of the seed generator is to produce a representation that incorporates coordinates and features for a low-resolution, fully defined sketch point cloud, referred to as the ``seed." To leverage the powerful detail retention capabilities of the Upsample Transformer~\cite{zhou2022seedformer}, the inputs \textbf{f}, $F_{p}$, and $P_{p}$ are fed into the Upsample Transformer, which then generates point features. This process helps capture the present and missing shape details through point-level segmentation operations. Subsequently, each generated point feature is combined with the shape code using a MLP to create a coarse point cloud, denoted as $P_{c}\in \mathbb{R}^{N_{c}\times 3}$. The generated coarse point cloud is then concatenated with the original input point cloud $P\in \mathbb{R}^{N \times 3}$, and this merged point cloud is downsampled using FPS~\cite{qi2017pointnet++} to yield $P_{0}\in \mathbb{R}^{N_{0}\times 3}$. Here, $N_{c}$ denotes the number of points in the coarse point cloud, while $N_{0}$ refers to the number of points after downsampling the merged cloud. Finally, the resulting downsampled point cloud, $P_{0}\in \mathbb{R}^{N_{0}\times 3}$, is used as the seed point cloud for the point generation module.

\textbf{Point generation.} As shown in Figure~\ref{fig:2} (b). The point generation module
consists of three steps of DFG; each block has PointNet~\cite{qi2017pointnet}, Graph Construction Module , Graph Aggregation Module, Graph Fusion Module, MLP, and Deconvolution~\cite{wen2021pmp}. Given a rough point cloud $P_{i-1} \in \mathbb{R}^{N_{i-1}\times C}$ and \textbf{f}, we first adopt a PointNet~\cite{qi2017pointnet} to learn per-point features $Q _{i-1} \in \mathbb{R} ^{N_{i-1} \times C}$ . Then, $Q _{i-1} \in \mathbb{R} ^{N_{i-1} \times C}$ and the feature $H_{i-1} \in \mathbb{R} ^{N_{i-1} \times C}$ passed from the previous module are used as the inputs for the graph construction to obtain graph $G_{i}$. Next, $G_{i}$ , $Q _{i-1}$, and $H_{i-1}$ are fed into graph aggregation module and graph fusion module to capture detail-rich geometric features and obtain feature $H_{i}$. Finally, we first duplicate the input point set $P_{i-1} \in \mathbb{R} ^{N_{i-1} \times 3}$ to obtain an intermediate set $\hat{P_{i} }\in \mathbb{R} ^{N_{i} \times 3}$. Then, we employ a MLP and Deconvolution to predict the offsets $\bigtriangleup P_{i}$ for each point, enabling us to obtain the upsampled coordinates  $P_{i} \in \mathbb{R} ^{N_{i} \times 3}$.

\subsection{Graph Construction}
Traditional point cloud completion techniques often use pooling operations to condense partial input observations into a unified global feature vector. This vector is then used during the decoding stage to reconstruct the shape. However, these pooling operations can cause the loss of crucial geometric details, which reduces the effectiveness of global features for completing the point cloud. To resolve this, we introduce a flexible degree point graph that operates across both local and global scales within the DFG model. This method fully capitalizes on the dynamic nature of the graph structure, thereby improving the overall performance of point cloud completion.

\begin{figure}[t]
	\centering
	\includegraphics[width=1\columnwidth]{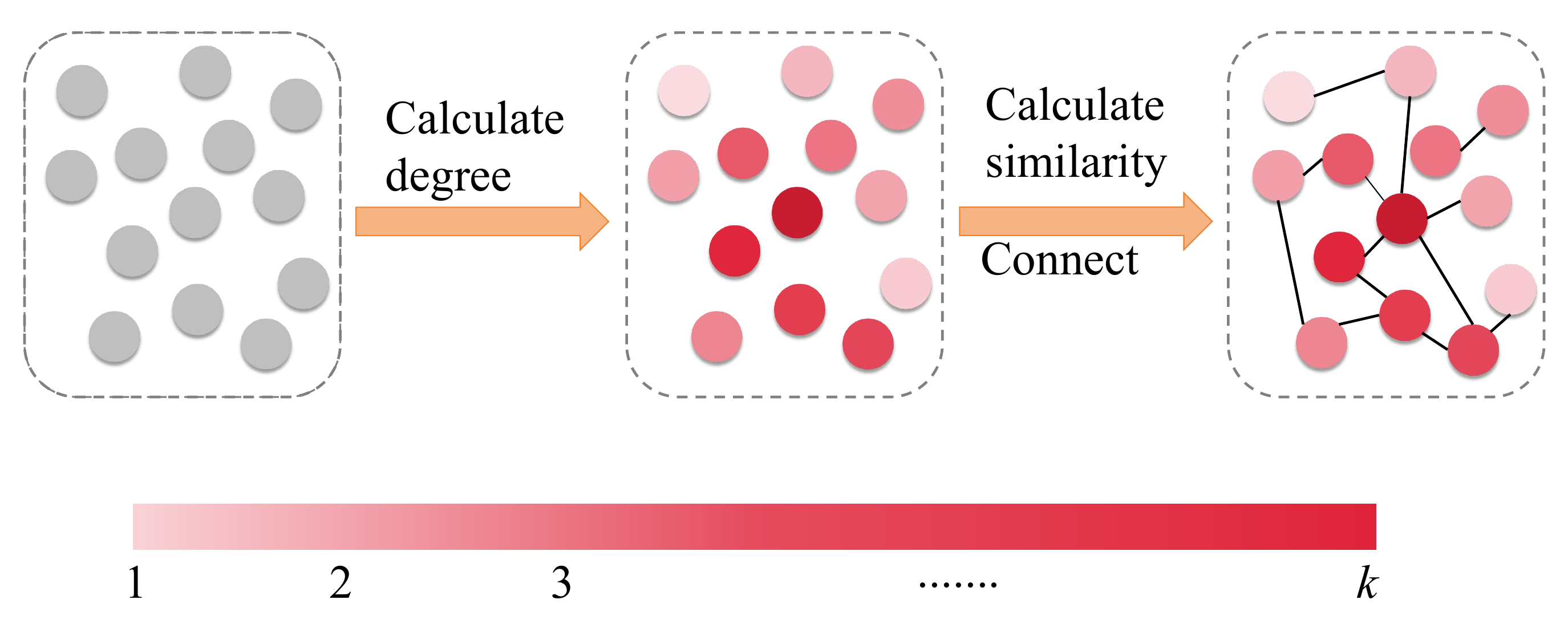}
	\caption{The detailed structure of graph construction. The grey balls are point features, and the red color's depth represents the graph degree's size. The deeper the red, the higher the degree of the graph.}
	\label{fig:3}
\end{figure}

\begin{figure}[t]
	\centering
	\includegraphics[width=1\columnwidth]{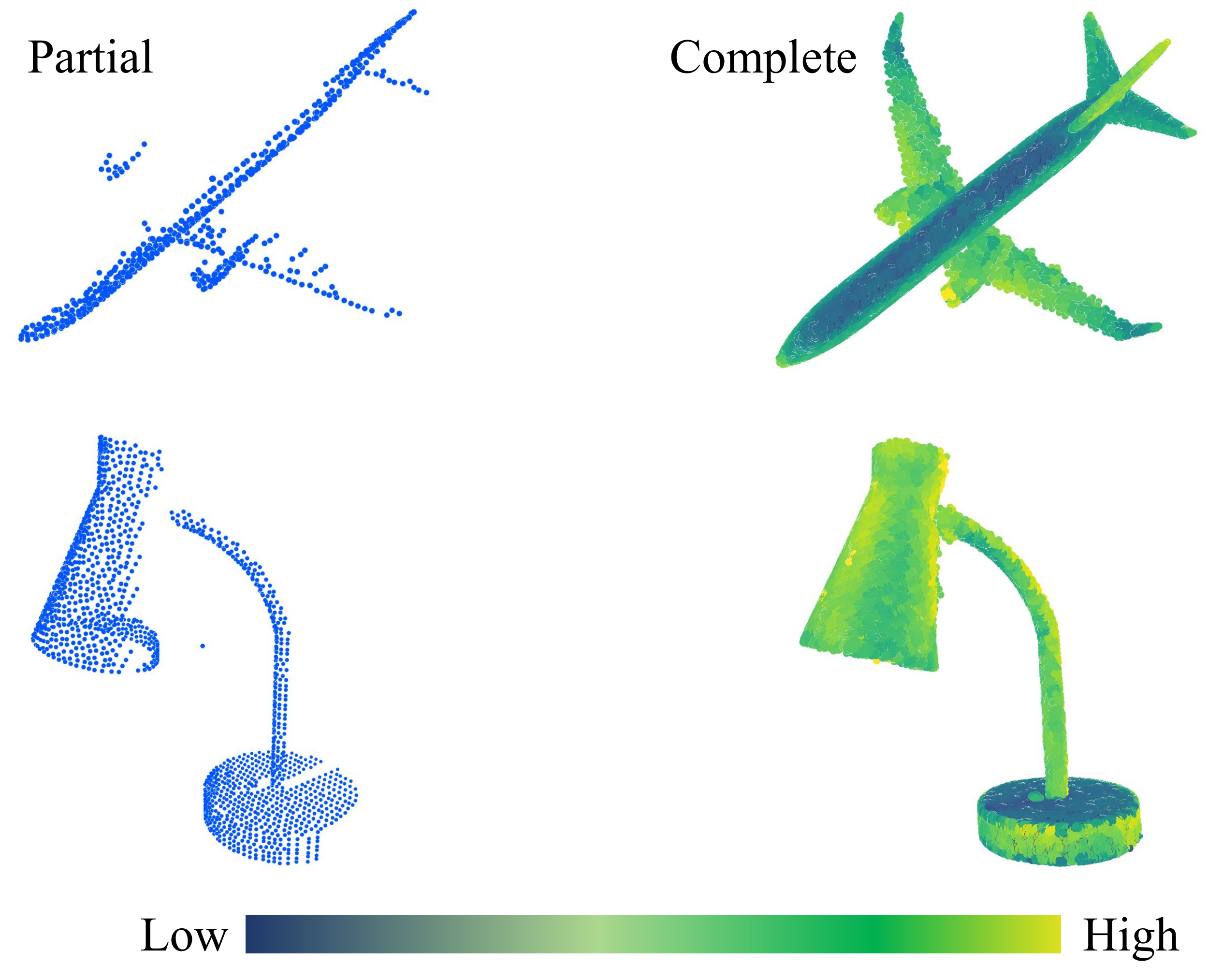}
	\caption{Visualization of the detail richness metric. The left side displays the incomplete input point cloud, and the right side shows the corresponding complete reconstruction.}
	\label{fig:4}
\end{figure}

\textbf{Degree Flexibility.} Figure~\ref{fig:3} illustrates the specific operations of the graph construction module. We assign different node degrees to points based on a detail-rich metric, which is used to identify points that require more reconstruction effort. Specifically, we denote a set of parent points $P_{i-1} \in \mathbb{R} ^{N_{i-1} \times 3}$ obtained from the previous step. First, we extract the per-point features $Q _{i-1} \in \mathbb{R} ^{N_{i-1} \times C}$ from the shape code \textbf{f} using the basic PointNet framework. These features $Q _{i-1} \in \mathbb{R} ^{N_{i-1} \times C}$ are then fed into the graph construction module for point cloud density allocation. In the graph construction module, the downsampling ratio $s$ is set to be 3 to prevent significant loss of geometric information. We initially utilize Farthest Point Sampling (FPS)~\cite{qi2017pointnet++} to perform downsampling on the point cloud, yielding a set of downsampled points $P_{d} \in \mathbb{R} ^{N_{d} \times 3}$ and their corresponding features $Q_{d} \in \mathbb{R} ^{N_{d} \times C}$, where $N_{d}$ indicates the number of points in the downsampled cloud. Subsequently, we perform interpolation upsampling on the downsampled points to retrieve the points $P_{u} \in \mathbb{R} ^{N \times 3}$ and corresponding features $Q_{u} \in \mathbb{R} ^{N \times C}$. $H_{i-1}$ represents the features extracted from the previous layer. The detail richness metric for each point $D_{Q} \in \mathbb{R} ^{N}$ is defined as the absolute difference between the feature maps after downsampling~$\downarrow$  and upsampling~$\uparrow$ and the original feature map:
\begin{equation}\label{eq:1} 
	D_{Q}=\sum_{C} \left | Q-Q_{\downarrow \uparrow }  \right |+\left | Q-H_{i-1}  \right |. 
\end{equation}
The overall degree budget is based on the degree assigned to each point in the feature map, which is proportional to its corresponding point value. To improve the sensitivity of the graph structure to geometric variations, we integrate curvature into the degree computation. While the detail richness metric reflects variations in feature space, curvature captures local geometric complexity. By combining these two factors, the model can more effectively identify and allocate more connections to structurally important points, thereby improving overall reconstruction quality. 

To further illustrate the effectiveness of the proposed detail richness metric, we visualize its spatial distribution in Figure~\ref{fig:4}. As shown, regions with complex geometry or large missing parts—such as the airplane's wingtips and the lamp's edges—are assigned higher detail-richness scores. In contrast, flat and complete regions receive lower scores. This distribution demonstrates that the proposed metric effectively highlights structurally important areas, guiding adaptive degree allocation and feature fusion in point cloud completion.

The degree $d_{i}$ assigned to each point is computed as follows:
\begin{subequations}\label{eq:2} 
	\begin{align}
		d_i = & \mathrm{round} \left( B \times \frac{D_i}{\sum_{i=1}^{n} D_i} \right), \\
		B &= \alpha \sum_{i=1}^{n} (D_i + \kappa_i). 
	\end{align}
\end{subequations}

\begin{figure}[t]
	\centering
	\includegraphics[width=1\columnwidth]{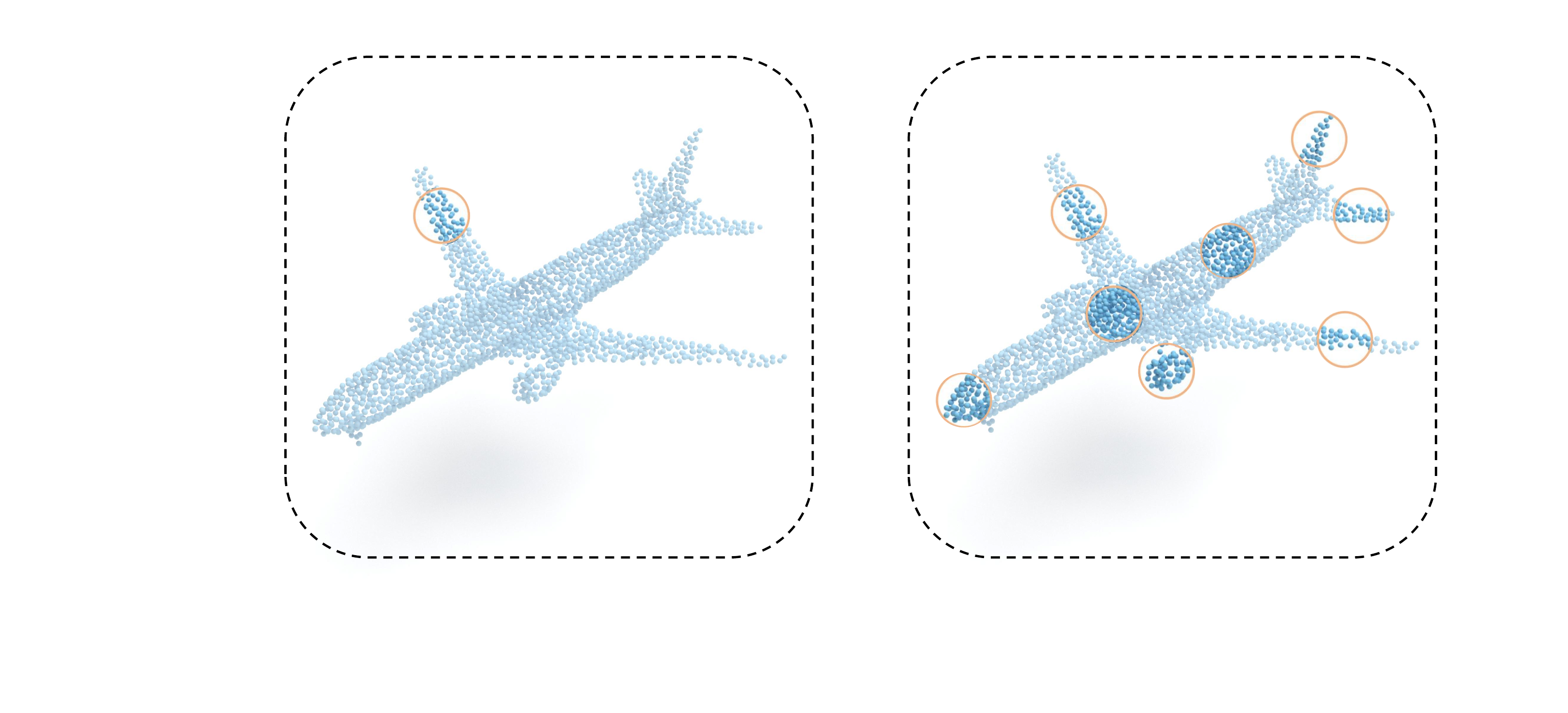}
	\caption{An illustration of local graph (left) and global graph (right) sampling strategies. In the local graph (left), the circled points represent  nodes selected from local neighborhoods to capture fine-grained details. In the global graph (right), the circled points are sampled from the entire point cloud to establish long-range contextual connections.}
	\label{fig:5}
\end{figure}

Here, \textit{round}$(\cdot)$ denotes standard rounding to the nearest integer, $D_{i}$ represents the detail richness metric of point, $n$ represents the number of points in the point cloud within the region, $\alpha$ is a scaling hyperparameter that controls the overall connection budget, $\kappa_{i}$ denotes the curvature of point, The total degree budget $B$ is adaptively computed by incorporating both the detail metric and curvature, ensuring that points in high-detail or highly curved regions receive more connections, thereby enhancing reconstruction precision in those areas.

\textbf{Space Flexibility.} We enhance the spatial flexibility of the DFG through effective searching of connections between local and global point nodes. Both local and global information are crucial for point cloud processing. While the missing parts of the point cloud can be reconstructed from local regions, they can also learn from distant areas with similar features for further refinement.

Existing graph-based algorithms have demonstrated some degree of spatial flexibility. Unlike convolutional and window attention mechanisms that consistently focus on fixed neighboring regions, graph aggregation can flexibly attend to crucial parts of the graph without spatial constraints. However, current graph-based methods still need to improve spatial flexibility, primarily in their inability to simultaneously integrate local and global information.

\begin{figure*}
	\centering
	\includegraphics[width=1.5\columnwidth]{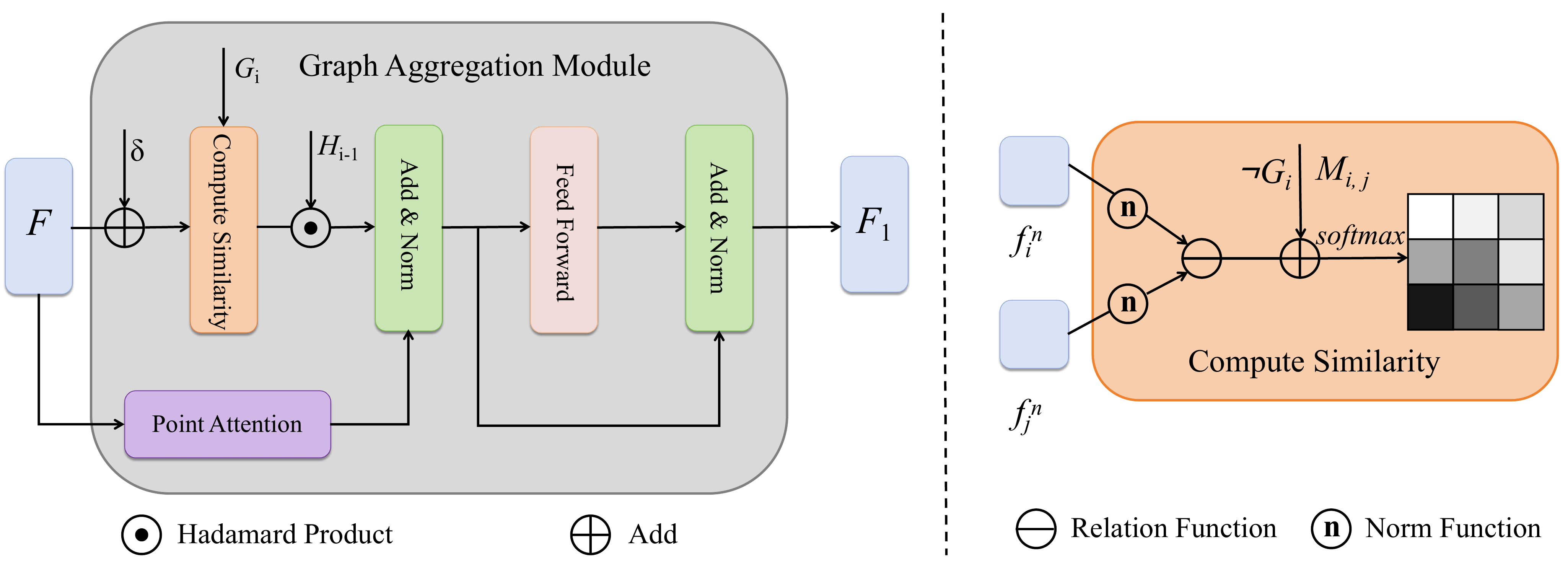}
	\caption{The detailed structure of the Graph Aggregation Module (GAM). Here $\delta$ is the positional encoding, $\neg$ is the logical inversion symbol.}
	\label{fig:6}
\end{figure*}

We aim to capture local and global features to enhance efficiency in point cloud completion. Local features contribute to detailed reconstruction, while global features are derived from distant points. To achieve this balance, we propose two sampling methods. These techniques effectively gather information from local neighborhoods and the overall point cloud. For local graph construction, we adopt a fixed neighborhood size of $k=16$, following standard practice in prior point cloud learning works such as PointNet++~\cite{qi2017pointnet++}, which has been widely validated for balancing local geometric representation and computational efficiency. For the global graph, we sample a fixed number of 512 points using FPS to ensure broad spatial coverage across the entire point cloud, enabling long-range contextual information to be captured effectively. As illustrated in Figure~\ref{fig:5}, local sampling focuses on densely connected regions to preserve fine-grained details, while global sampling captures semantically relevant but spatially distant regions to enhance structural coherence.

\subsection{Graph Feature Integration}
Figure~\ref{fig:6} shows the structure of Graph Aggregation. After constructing the flexible graph structure, we proceed with graph aggregation, enabling each node to communicate effectively with its connected neighbors and utilize their information for self-optimization, thereby enhancing point cloud completion. In the application of graphs to point clouds, aggregation methods such as max pooling or edge-conditioned forms are typically preferred. We favor edge-conditioned aggregation because max pooling can significantly lose vital information from neighboring points, which is crucial for point cloud tasks. To improve the geometric sensitivity of the graph aggregation process, we incorporate the Manhattan distance into the computation of edge weights. Unlike Euclidean distance, which treats all spatial directions uniformly, Manhattan distance exhibits greater sensitivity to axis-aligned variations, making it more effective for capturing directional structures, boundary transitions, and geometric discontinuities. This choice is motivated by prior work in point cloud understanding~\cite{xiang2021walk}, where a relative positional encoding based on Manhattan distance was shown to enhance the representation of local geometric relationships. Inspired by this insight, we integrate Manhattan distance into our edge-conditioned aggregation framework to better capture fine-grained structural cues. Since point cloud completion relies heavily on preserving detailed local geometry, adopting such a direction-sensitive distance metric enables more informative feature propagation. In addition, edge-conditioned aggregation inherently emphasizes feature interactions between connected nodes, allowing our approach to retain richer neighborhood information critical for accurate reconstruction.

\begin{figure}
	\centering
	\includegraphics[width=1\columnwidth]{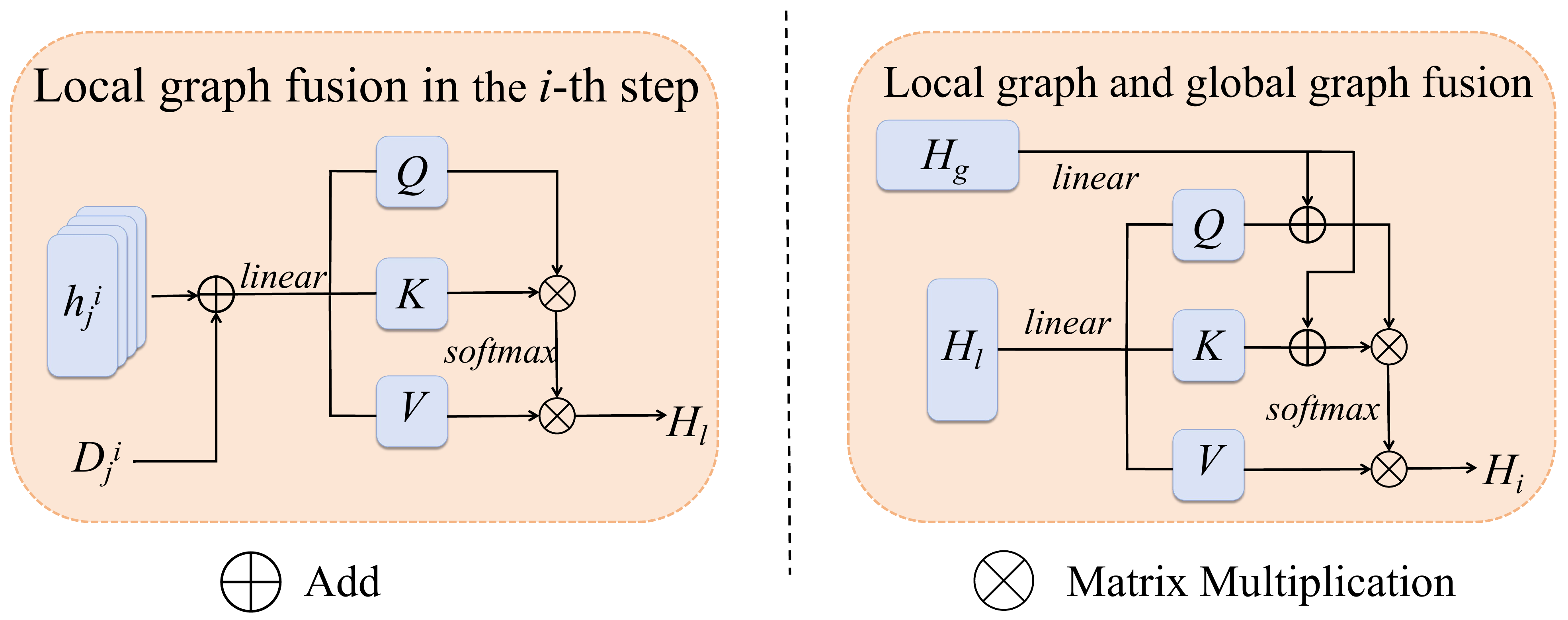}
	\caption{Structure of the Graph Fusion Module (GFM), which integrates local and global graph features using attention, guided by detail richness metric.}
	\label{fig:7}
\end{figure}

The mathematical formulation for edge-conditioned aggregation in the $i$-th layer of DFG is as follows: Given the node features $f_{j} ^{i}\in F_{i}$ as input and the set of adjacent nodes $\mathcal{N}\left ( j \right )$, the output for node $j$ is computed as $h_{j}^{i}$:
\begin{subequations}\label{eq:3} 
	\begin{align}
		\hat{a} ^{i} _{j,k}& =\beta_{i}  \left (( f_{j} ^{i} +\delta  )\ominus ( f_{k} ^{i } +\delta )+C\times (\neg G_{i})+{M} ^{i} _{j,k} \right ), \\
		&{M} ^{i} _{j,k}=\gamma \left (\left | x_{j} ^{i} - x_{k} ^{i}\right | + \left | y_{j} ^{i} - y_{k} ^{i}\right |+\left | z_{j} ^{i} - z_{k} ^{i}\right |\right ). 
	\end{align}
	
\end{subequations}

\noindent where $\beta_{i}$ is a linear projection function implemented with MLP, $\delta$ is a positional encoding vector to enhance spatial relation information, $\gamma$ represents a hyperparameter, $\ominus$ is a parameterized function that measures the similarity between node pairs $(i,j)$,  $C $ is a negative constant, and $\neg G_{i}$ represents the logical negation operation on the graph, indicating nodes that are not connected in the graph, ${M} ^{i} _{j,k}$ represents the Manhattan distance between two points. Additionally, to normalize the computed weights for balancing the scale, $\hat{a} ^{i} _{j,k}$ is applied using the softmax function:
\begin{equation}\label{eq:4}
	a^{i} _{j,k} =\frac{\mathrm {exp}\left ( \hat{a} ^{i} _{j,k} \right ) }{\sum_{k\in \mathcal{N}(j)}^{} \mathrm {exp}\left ( \hat{a} ^{i} _{j,k} \right ) }.      
\end{equation}

Finally, the features after graph aggregation can be obtained as follows:
\begin{equation}\label{eq:5}
	h^{i} _{j}= \sum_{k\in \mathcal{N}(j)}^{} a^{i} _{j,k} \odot \left ( h^{i-1} _{k}+\delta \right ),
\end{equation}
where, $\odot$ is Hadamard product, and $h^{i-1} _{k}$ is from the previous module after graph aggregation.

However, due to the significant variation in detail richness across different point cloud regions, relying on a single-scale aggregation strategy remains insufficient for modeling complex geometric structures. To address this issue, we introduce the Graph Fusion Module (GFM), as illustrated in Figure \ref{fig:7}, which further integrates features from both local and global graphs to enhance overall representation capability.

Specifically, for the features within each region, guided by the detail richness metric of the region, we first transform these features into query, key, and value tokens through linear projection. Subsequently, we compute self-attention weights among regions and normalize these weights using a softmax function to achieve adaptive weighted integration of regional features, thereby emphasizing those critical to detail reconstruction. Additionally, to leverage the complementary nature of multi-scale information, we further fuse the local and global scale features through a linear transformation after processing by the two-scale attention modules. This fusion effectively enhances the integrity of feature representation, ensuring that the model simultaneously preserves the precision of local structures and the consistency of global structures.

\begin{subequations}\label{eq:8}
	\begin{align}
	\hat{{H}_{l}}&=\left ( h_{j} ^{i}+ D_{j} ^{i} \right ),\\
	Q=\hat{{H}_{l}} W_{Q}, \quad &K=\hat{{H}_{l}}W_{K},\quad V=\hat{{H}_{l}}W_{V}.\\
	{H}_{l}=&\mathrm{softmax}(\frac{QK^{T} }{\sqrt{d_{k}}})V.
		\end{align}
\end{subequations}

\begin{subequations}\label{eq:8}
	\begin{align}
		Q&=H_{l} W_{Q},\quad K=H_{l} W_{K},\quad V=H_{l} W_{V},\\
		{H}_{i}&=\mathrm{softmax}(\frac{(Q+\hat{H_{g} })(K+\hat{H_{g} })^{T} }{\sqrt{d_{k} } })V.
	\end{align}
\end{subequations}
Where ${H}_{l}$ denotes the output feature of the local graph fusion, $h_{j} ^{i}$ represents the feature of local graph at layer $i$, $D_{j} ^{i}$ represents detail richness of the region, $W_{Q}, W_{K}, W_{V} $ are learnable projection matrices for queries, keys, and values, respectively. $Q, K, V$ denote the projected query, key, and value embeddings. The scalar $d_{k}$ is the dimensionality of the key vector and is used for scaling during attention computation. $\hat{H_{g} }$ represents the global graph feature after linear transformation.

\subsection{Training Loss}
For training, we use Chamfer Distance (CD)~\cite{fan2017point} as a metric to compute the point cloud loss. Given the predicted set of point clouds $\left \{P_{0}, P_{1}, P_{2}, P_{3}\right \}$ and the corresponding ground truth set $\left \{S_{0},S_{1},S_{2},S_{3}\right \}$ , both generated through FPS, the model's loss function can be formulated as follows:
\begin{equation}\label{eq:6}
  \mathcal{L} =\sum_{i=0}^{} CD(P_{i},S_{i}),
\end{equation}
The term CD refers to the loss of chamfer distance, which can be defined as follows:
\begin{equation}\label{eq:7}
	CD(\mathcal{P},\mathcal{S})=\frac{1}{\left | \mathcal{P}  \right | } \sum_{p\in \mathcal{P}}^{}
	\min_{s\in\mathcal{S}} \left \| p-s \right \| +\frac{1}{\left | \mathcal{S}  \right | } \sum_{s\in \mathcal{S}}^{}
	\min_{p\in\mathcal{P}} \left \| s-p \right \|.
\end{equation}

\begin{table*}
	\centering
	\renewcommand\arraystretch{1.2}
	\begin{threeparttable}
		\caption{The results of the PCN dataset indicate the $\ell_{1}$ Chamfer Distance multiplied by $10^3$, with lower values reflecting better performance. The best results are highlighted in bold.}
		\label{tb:1}		
		\begin{tabularx}{\textwidth}{c|>{\centering\arraybackslash}X|>{\centering\arraybackslash}X>{\centering\arraybackslash}X>{\centering\arraybackslash}X>{\centering\arraybackslash}X>{\centering\arraybackslash}X>{\centering\arraybackslash}X>{\centering\arraybackslash}X>{\centering\arraybackslash}X>{\centering\arraybackslash}X}
			\toprule
			Methods  & Average			  & Plane 				& Cabinet 			  & Car 	   & Chair 	& Lamp 	 & Couch   & Table    			& Boat      \\
			\midrule
			FoldingNet~\cite{yang2018foldingnet}    & 14.31          & 9.49 & 15.80 		  & 12.61 & 15.55 & 16.41 & 15.97 & 13.65 			& 14.99  \\
			PCN ~\cite{yuan2018pcn}  & 9.64 & 5.50          & 22.70 	      & 10.63 & 8.70 & 11.00 & 11.37 & 11.68 			& 8.59  \\
			TopNet~\cite{tchapmi2019topnet}         & 12.15 & 7.61          & 13.31 		  & 10.90 & 13.82 & 14.44 & 14.78 & 11.22 			& 11.12  \\
			AtlasNet~\cite{groueix2018papier}     & 10.85          & 6.37 & 11.94		  & 10.10 & 12.06 & 12.37 & 12.99 & 10.33 			& 10.61  \\
			GRNet~\cite{xie2020grnet}        & 8.83          & 6.45 & 10.37 		  & 9.45 & 9.41 & 7.96 & 10.51 & 8.44 			& 8.04  \\	
			PoinTr~\cite{yu2021pointr}       & 8.38 		  & 4.75          & 10.47 		  & 8.68 & 9.39 & 7.75 & 10.93 & 7.78 & 7.29  \\
			CRN~\cite{wang2020cascaded}       & 8.51          & 4.79 			& 9.97 & 8.31 & 9.49 & 8.94 & 10.69 & 7.81 			& 8.05  \\
			NSFA~\cite{zhang2020detail}         & 8.06 & 4.76 			& 10.18 		  & 8.63 & 8.53 & 7.03 & 10.53 & 7.35		& 7.48  \\			
			SnowFlake~\cite{xiang2021snowflakenet}        & 7.21 & 4.29			& 9.16 		  & 8.08 & 7.89 & 6.07 & 9.23 & 6.55 			& 6.40  \\
			FBNet~\cite{yan2022fbnet} & 6.94 & 3.99 			& 9.05 		  & 7.90 & 7.38 & 5.82 & 8.85 & 6.35 			& 6.18  \\
			SeedFormer~\cite{zhou2022seedformer}       & 6.74 & 3.85 			& 9.05 		  & 8.06 & 7.06 & 5.21 & 8.85 & 6.05 			& 5.85  \\
			AnchorFormer~\cite{chen2023anchorformer}        & 6.59		  & 3.70 			& 8.94 		  & 7.57 & 7.05 & 5.21 & 8.40 & 6.03 & 5.81  \\
			POINTATTN~\cite{wang2024pointattn}       & 6.86          & 3.87 			& 9.00 & 7.63 & 7.43 & 5.90 & 8.68 & 6.32 			& 6.09  \\
			AdaPoinTr~\cite{10232862}    & 6.53          & 3.68 			& 8.82 & 7.47 & 6.85 & 5.47 & \textbf{8.35} & 5.80		& 5.76 \\			
			
			\midrule
			Ours     & \textbf{6.42} &\textbf{3.62} & \textbf{8.63} 		  & \textbf{7.56} & \textbf{6.71} & \textbf{4.99} & 8.41 & \textbf{5.78} 			& \textbf{5.68}  \\
			\bottomrule
		\end{tabularx}
	\end{threeparttable}
\end{table*}

\begin{figure*}[t]
	\centering
	\includegraphics[width=2\columnwidth]{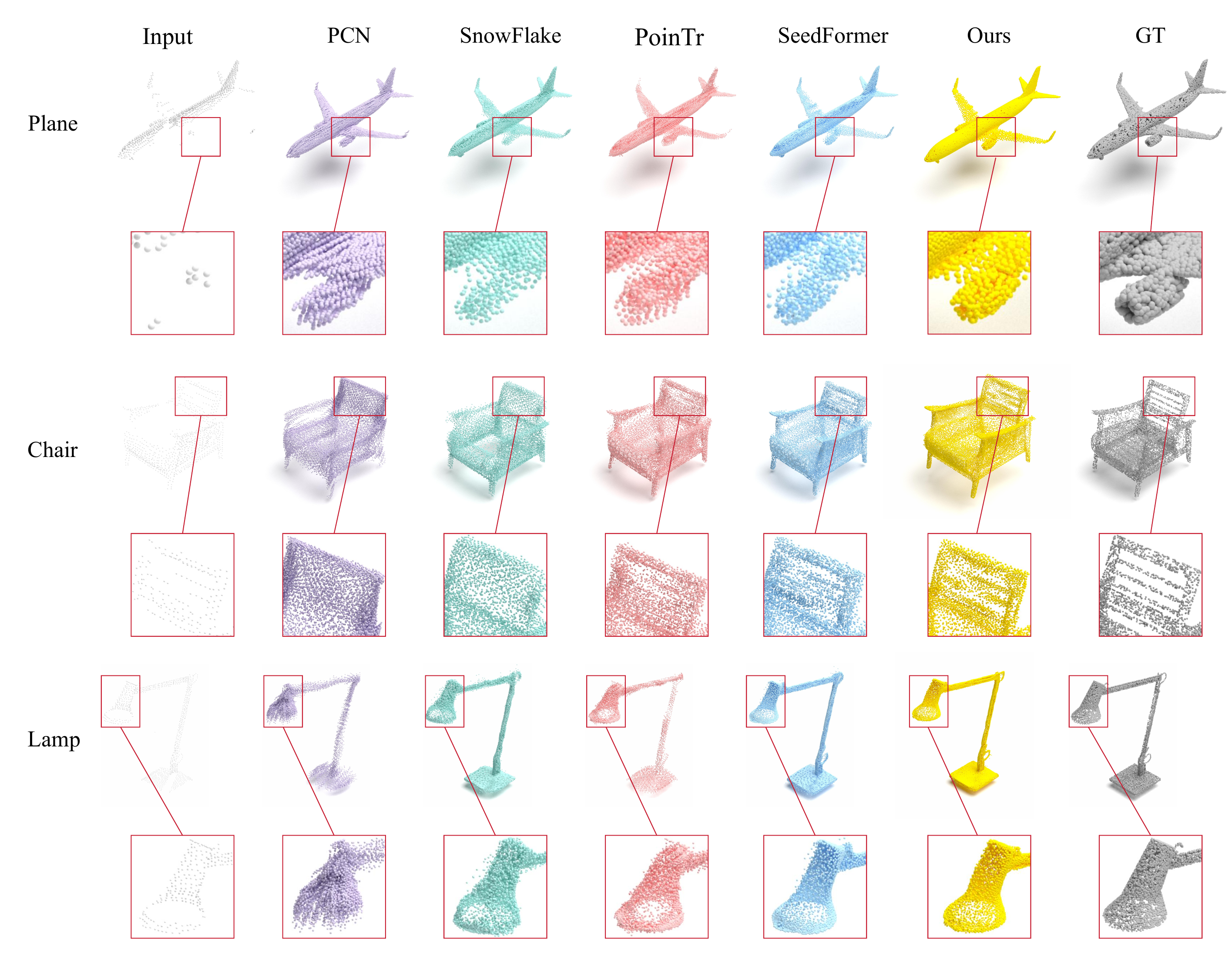}
	\caption{ Three visual examples of point cloud completion outcomes from various PCN dataset methods are shown. Different colors represent the point clouds reconstructed by each approach.
			}
	\label{fig:8}
\end{figure*}

\section{Experiments}\label{exp}
To thoroughly validate the effectiveness of our DFG-PCN, we perform extensive experiments on four widely recognized benchmarks: PCN~\cite{yuan2018pcn}, ShapeNet-55, ShapeNet-34~\cite{yu2021pointr} and KITTI~\cite{geiger2013vision}. The results demonstrate that our method is more effective than the current state-of-the-art techniques in point cloud completion.
\subsection{Dataset and Evaluation Metric}
\textbf{The PCN Dataset.} The PCN dataset, as introduced by Yuan et al.~\cite{chang2015shapenet}. It includes paired data sets of partial and complete point clouds, making it ideal for point cloud completion tasks. To ensure consistency with prior work~\cite{yuan2018pcn,xiang2021snowflakenet,zhou2022seedformer}, we adopt the same approach for data partitioning during training and evaluation. Specifically, the dataset is split into 28,974 training samples, 800 validation samples, and 1,200 test samples. Additionally, to normalize the point cloud sizes, we resample all incomplete point clouds to 2,048 points, addressing variability in point counts while aligning with standard practices. This ensures uniformity across experiments, enabling reliable comparison and evaluation.

\textbf{ShapeNet-55 Dataset.} The ShapeNet-55 dataset~\cite{yu2021pointr} is a large-scale repository of 3D object models constructed based on the ShapeNet dataset~\cite{chang2015shapenet}. It encompasses a diverse range of everyday object categories, making it a comprehensive resource for 3D shape analysis tasks. The ShapeNet-55 dataset contains 3D shapes across 55 categories, with 41,952 models allocated for training and 10,518 models for testing. In our experiments conducted on the ShapeNet-55 benchmark, we adhere to the experimental setup of PoinTr~\cite{yu2021pointr}, ensuring that the evaluation process is consistent and comparable with prior research. This adherence allows for an objective and fair performance comparison across different models and approaches within the same framework.

\textbf{ShapeNet-34 Dataset.} The ShapeNet-34 dataset~\cite{yu2021pointr} comprises 46,765 3D objects spanning 34 different categories. The test set is divided into two parts: 3,400 objects from the 34 visible categories and 2,305 objects from 21 unseen categories. This allows for evaluating both within-category and cross-category generalization. In line with previous research~\cite{yu2021pointr,10232862,chen2023anchorformer}, we evaluate models on point cloud data with varying levels of incompleteness, specifically where 25\%, 50\%, and 75\% of points are missing. These levels of missing data represent three different degrees of difficulty for the completion task: easy (S), medium (M), and hard (H), respectively, offering a comprehensive evaluation across different scenarios.

\textbf{KITTI Dataset.} To assess the effectiveness of the proposed model on real-world scanned data, we conduct additional experiments using the KITTI~\cite{geiger2013vision} dataset, which comprises sequences of LiDAR scans captured from outdoor environments. Car instances are isolated in each frame based on their 3D bounding boxes, resulting in 2,401 partial point clouds. Unlike synthetic datasets, KITTI features real-world data that is often highly sparse and lacks complete ground-truth point clouds, presenting greater challenges for point cloud completion.
Following the standard evaluation protocol adopted in~\cite{zhou2022seedformer,chen2023anchorformer}, all car shapes in the KITTI dataset are used solely for testing. The models are trained on the car subset of the PCN dataset~\cite{yuan2018pcn}, which contains complete car shapes. This setup verifies the model's ability to generalize from synthetic training data to sparse, incomplete real-world scans.

\begin{table*}
	\centering
	\renewcommand\arraystretch{1.2}
	\begin{threeparttable}
		\caption{The results of the ShapeNet-55 dataset indicate the $\ell_{2}$ Chamfer Distance multiplied by $10^3$, with lower values reflecting better performance. The best results are highlighted in bold.}
		\label{tb:2}
		\begin{tabularx}{\textwidth}{c
				|cc>{\centering\arraybackslash}X>{\centering\arraybackslash}cc
				|>{\centering\arraybackslash}X>{\centering\arraybackslash}cc>{\centering\arraybackslash}X>{\centering\arraybackslash}c|ccc|c}
			\toprule
			\multirow{2}{*}{Methods}  & \multirow{2}{*}{Table} 			  & \multirow{2}{*}{Chair} 				& Air plane 				& \multirow{2}{*}{Car} 	 	   	  & \multirow{2}{*}{Sofa} 	   & Bird house	  	& \multirow{2}{*}{Bag}   & \multirow{2}{*}{Remote}    & Key board &\multirow{2}{*}{Rocket} & \multirow{2}{*}{CD-S} & \multirow{2}{*}{CD-M} &\multirow{2}{*}{CD-H} & \multirow{2}{*}{CD-Avg}      \\
			\midrule	
			FoldingNet~\cite{yang2018foldingnet}    & 2.53          & 2.81 & 1.43 		  & 1.98 & 2.48 & 4.71 & 2.79 & 1.44 	& 1.24 & 1.48 & 2.67 & 2.66 & 4.05 & 3.12 \\
			PFNet~\cite{huang2020pf} & 3.95 & 4.24 & 1.81 & 2.53 & 3.34 & 6.21 & 4.96 & 2.91 & 1.29 & 2.36 & 3.83 & 3.87 & 7.97 & 5.22 \\
			TopNet~\cite{tchapmi2019topnet}    & 2.21          & 2.53 & 1.14 		  & 2.28 & 2.36 & 4.83 & 2.93 & 1.49 	& 0.95 & 1.32 & 2.26 & 2.16 & 4.30 & 2.91 \\
			PCN~\cite{yuan2018pcn}    & 2.13          & 2.29 & 1.02 		  & 1.85 & 2.06 & 4.50 & 2.86 & 1.33 	& 0.89 & 1.32 & 1.94 & 1.96 & 4.08 & 2.66 \\
			GRNet~\cite{xie2020grnet}    & 1.63          & 1.88 & 1.02 		  & 1.64 & 1.72 & 2.97 & 2.06 & 1.09 	& 0.89 & 1.03 & 1.35 & 1.71 & 2.85 & 1.97 \\
			PoinTr~\cite{yu2021pointr}    & 0.81          & 0.95 & 0.44 		  & 0.91 & 0.79 & 1.86 & 0.93 & 0.53 	& 0.38 & 0.57 & 0.58 & 0.88 & 1.79 & 1.09 \\
			SnowFlake~\cite{xiang2021snowflakenet}    & 0.98          & 1.12 & 0.54 		  & - & - & 1.93 & 1.08 & - 	& 0.48 & - & 0.70 & 1.06 & 1.96 & 1.24 \\
			SeedFormer~\cite{zhou2022seedformer}    & 0.72          & 0.81 & 0.40		  & 0.89 & 0.71 & 1.51 &  0.79 & 0.46 	& 0.36 & 0.50 & 0.50 & 0.77 & 1.49 & 0.92  \\ 
			ODGNet~\cite{cai2024orthogonal}    & -          & - & -	 & - & - & - &  - & -	& -& - & 0.47 & 0.70 & 1.32 & 0.83 \\
			PPCL~\cite{fei2025point}    & 0.92          & 0.98 & 0.52	 & 0.94 & 0.87 & 1.56 &  0.98 & 0.51	& 0.41 & 0.50 & 0.61 & 0.95 & 1.89 & 1.15 \\
			 \midrule
			Ours    & \textbf{0.64}          & \textbf{0.77} & \textbf{0.34}		  & \textbf{0.85} & \textbf{0.64} & \textbf{1.42} &  \textbf{0.78} & \textbf{0.40}	& \textbf{0.35} & \textbf{0.48} & \textbf{0.45} & \textbf{0.67} & \textbf{1.26} & \textbf{0.79} \\
			\bottomrule
		\end{tabularx}
	\end{threeparttable}
\end{table*}

\begin{figure}
	\centering
	\includegraphics[width=1\columnwidth]{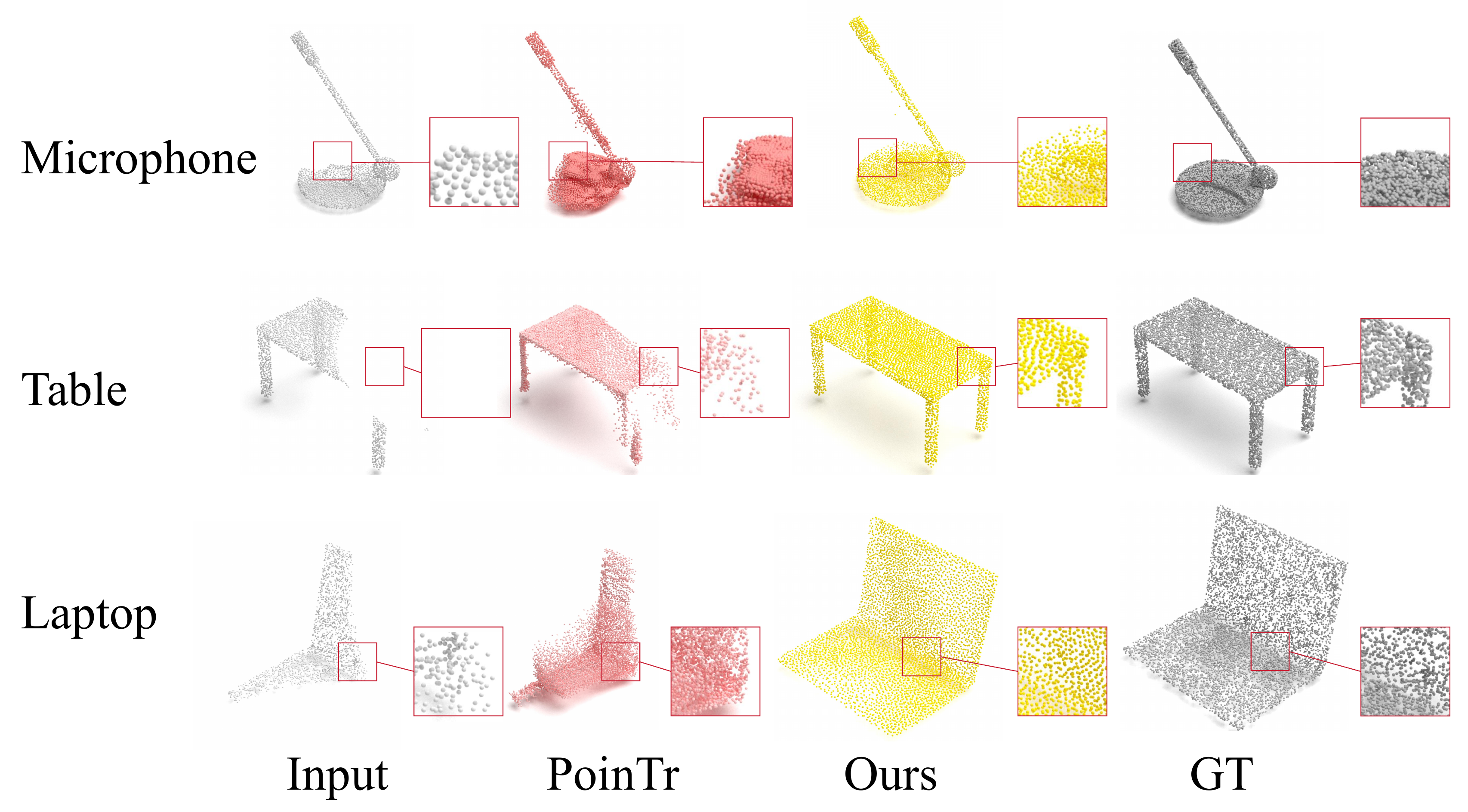}
	\caption{Qualitative results on ShapeNet-55. From top to bottom, 25\%, 50\%, and 75\% of the point cloud are masked, respectively.}
	\label{fig:9}
\end{figure}

\textbf{Evaluation metrics.} To assess the performance of different algorithms quantitatively, we employ two widely used metrics: CD$\text{-}\ell_{1}$ and CD$\text{-}\ell_{2}$. Both are based on CD, where lower values signify better performance, reflecting the closeness between the predicted and ground truth point clouds. For the KITTI benchmark, we adopt two quantitative metrics, Minimal Matching Distance (MMD) and Fidelity Distance (FD), to evaluate completion performance derived from the CD. MMD is computed using the CD and reflects the similarity between the completed output and a reference shape from ShapeNet, thereby evaluating how well the generated output aligns with typical object geometries. In contrast, FD quantifies the geometric fidelity of the completion by computing the average nearest-neighbor distance from each point in the input to the output, reflecting how accurately the model preserves the observed input structure.
These metrics provide a robust means to assess the accuracy and quality of 3D point cloud completion.

\textbf{Implementation Details.} The PyTorch framework is used to implement DFG-PCN. The model is trained on an NVIDIA GTX 4090 GPU with a batch size 32. We use the Adam optimizer~\cite{kingma2014adam}, with hyperparameters  $\beta1$ = 0.9 and $\beta2$ = 0.99. The learning rate is initialized at 0.001 and is reduced by a factor of 0.1 every 100 epochs. The training process spans 400 epochs, using the PCN dataset~\cite{yuan2018pcn} and the ShapeNet-55 and ShapeNet-34 datasets~\cite{yu2021pointr}.

\subsection{Results}

\begin{table*}
	\centering
	\renewcommand\arraystretch{1.2}
	\begin{threeparttable}
		\caption{The results of the ShapeNet-34 dataset indicate the $\ell_{2}$ Chamfer Distance multiplied by $10^3$, with lower values reflecting better performance. The best results are highlighted in bold.}
		\label{tb:3}
		\begin{tabularx}{\textwidth}{c|>{\centering\arraybackslash}X>{\centering\arraybackslash}X>{\centering\arraybackslash}X>{\centering\arraybackslash}X|>{\centering\arraybackslash}X>{\centering\arraybackslash}X>{\centering\arraybackslash}X>{\centering\arraybackslash}X}
			\toprule
			& \multicolumn{4}{c|}{34 seen categories} 		 &  \multicolumn{4}{c}{21 unseen categories}				  	             \\ 
			Methods&CD-S&CD-M&CD-H&CD-Avg&CD-S&CD-M&CD-H&CD-Avg\\
			\midrule	
			
			FoldingNet~\cite{yang2018foldingnet}    & 1.86        & 1.81 & 3.38 		  & 2.35 & 2.76 & 2.74 & 5.36 & 3.62 	\\
			TopNet ~\cite{tchapmi2019topnet}   & 1.77        & 1.61 & 3.54 		  & 2.31 & 2.62 & 2.43 & 5.44 & 3.50 	\\
			PCN~\cite{yuan2018pcn}    & 1.87        & 1.81 & 2.97 		  & 2.22 & 3.17 & 3.08 & 5.29 & 3.85 	\\
			PFNet~\cite{huang2020pf}    & 3.16        & 3.19 & 7.71 		  & 4.68 & 5.29 & 5.87 & 13.33 & 8.16 	\\
			GRNet~\cite{xie2020grnet}    & 1.26        & 1.39 & 2.57 		  & 1.74 & 1.85 & 2.25 & 4.87 & 2.99 	\\
			PoinTr~\cite{yu2021pointr}    & 0.76        & 1.05 & 1.88 		  & 1.23 & 1.04 & 1.67 & 3.44 & 2.05 	\\
			SnowFlake~\cite{xiang2021snowflakenet}    & 0.60        & 0.86 & 1.50 		  & 0.99 & 0.88 & 1.46 & 2.92 & 1.75 	\\
			ProxyFormer~\cite{li2023proxyformer}    & 0.44        & 0.67 & 1.33 		  & 0.81 & 0.60 & 1.13 & 2.54 & 1.42  \\
			SeedFormer~\cite{zhou2022seedformer}    & 0.48        & 0.70 & 1.30 		  & 0.83 & 0.61 & 1.07 & 2.35 & 1.34 \\
			PPCL~\cite{fei2025point}    & 0.62        & 0.95 & 1.87 		  & 1.15 & 0.81 & 1.35 & 2.92 & 1.69 \\  
			ODGNet~\cite{cai2024orthogonal}    & 0.44        & 0.64 & 1.14 		  & 0.75 & 0.59 & 1.01 & 2.26 & 1.29 \\
			\midrule
			Ours    & \textbf{0.42}        & \textbf{0.61} & \textbf{1.12} 		  & \textbf{0.72} & \textbf{0.57} & \textbf{0.96} & \textbf{2.15} & \textbf{1.22} \\
			\bottomrule
		\end{tabularx}
	\end{threeparttable}
\end{table*}

\begin{figure}
	\centering
	\includegraphics[width=1\columnwidth]{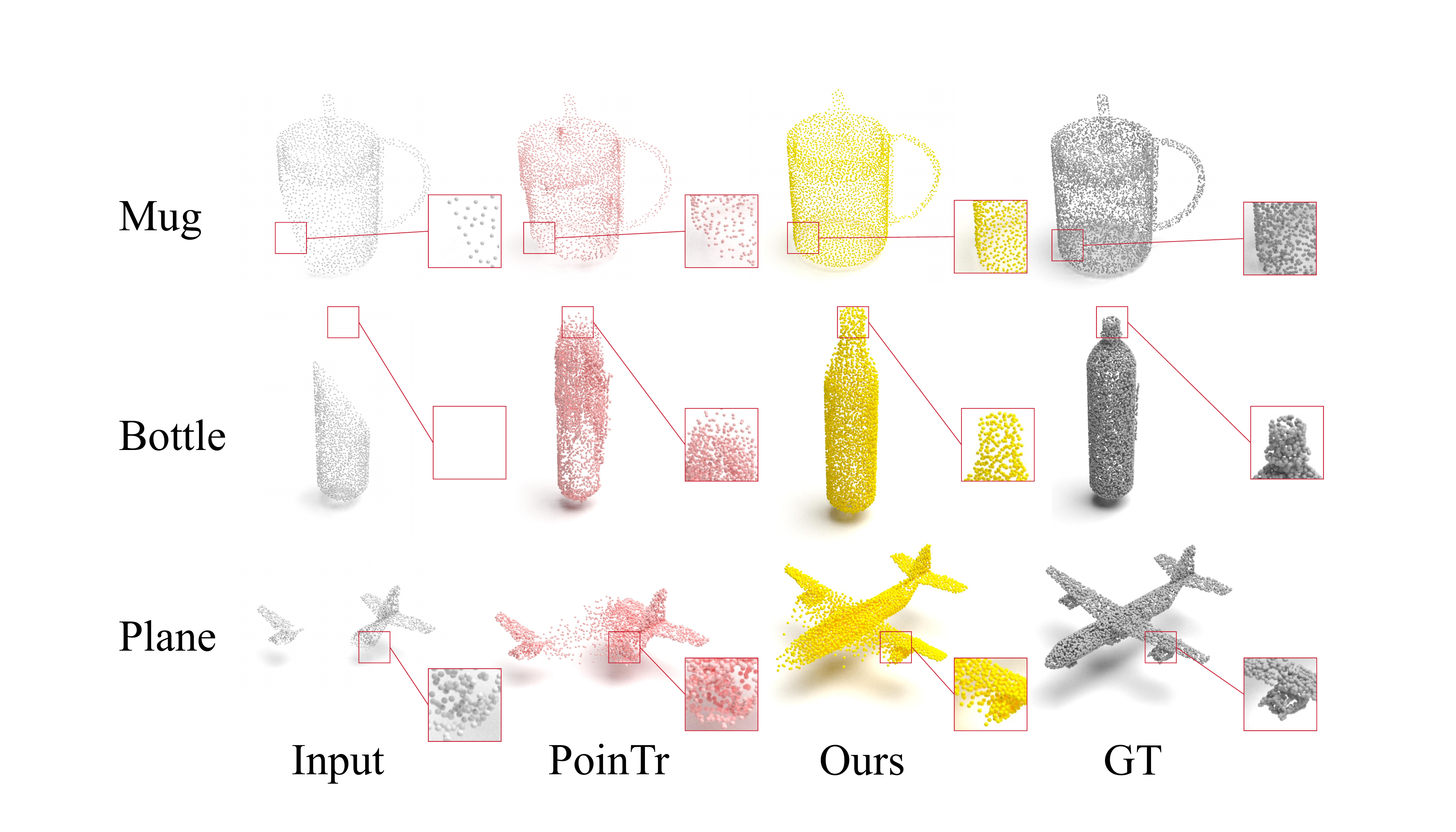}
	\caption{ Qualitative results on ShapeNet-34 seen categories. From top to bottom, 25\%, 50\%, and 75\% of the point cloud are masked, respectively.}
	\label{fig:10}
\end{figure}

\begin{figure}
	\centering
	\includegraphics[width=1\columnwidth]{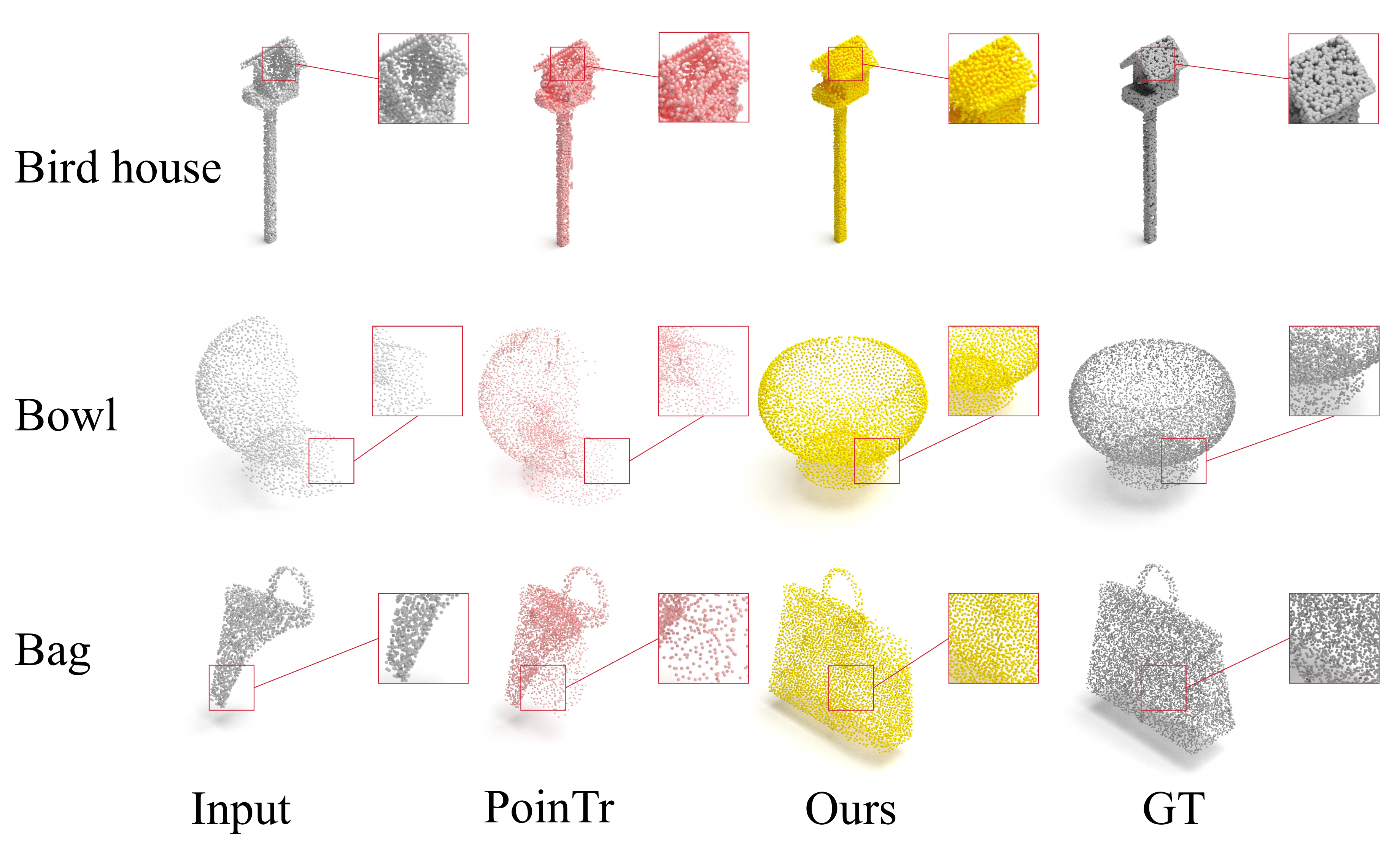}
	\caption{Qualitative results on ShapeNet-34 unseen categories. From top to bottom, 25\%, 50\%, and 75\% of the point cloud are masked, respectively.}
	\label{fig:11}
\end{figure}

\textbf{Evaluation on PCN Dataset.} Table \ref{tb:1} provides the quantitative evaluation of our method against other techniques across eight categories from the PCN dataset. Our method stands out with a notable average CD$\text{-}\ell_{1}$ score of 6.42, reflecting a clear improvement over existing approaches. Compared with the second-best method, AdaPoinTr~\cite{10232862}, our approach reduces 0.11 in the average CD, equating to a 1.68\% lower score (6.42 compared to 6.53). While some previous works in Table~\ref{tb:1} adopt similar coarse-to-fine strategies, DFG-PCN surpasses them by leveraging several novel design features. These results underscore the importance of learning robust feature representations for effective shape reconstruction.

Figure~\ref{fig:8} further visualizes three selected categories and directly compared the previous state-of-the-art methods. In particular, our method is capable of generating higher-quality object shapes, characterized by smoother surfaces (e.g., the body of an airplane) and more detailed local structures (e.g., the handle and lampshade of a lamp). Additionally, the point clouds produced by our method exhibit less noise. In contrast, other algorithms~\cite{zhou2022seedformer,yu2021pointr} may generate blurry shapes, often accompanied by outlier points scattered outside the main structure.

\textbf{Evaluation on ShapeNet-55.} We evaluate the effectiveness of DFG-PCN on the ShapeNet-55 dataset~\cite{yu2021pointr}, which features a more diverse set of categories. Table~\ref{tb:2} shows the $\ell_{2}$ Chamfer Distance (CD\text{-}$\ell_{2}$) results for different methods. Specifically, we report the average CD\text{-}$\ell_{2}$ scores across all categories, along with the CD\text{-}$\ell_{2}$ performance on point clouds with varying masking ratios, namely CD-S, CD-M, and CD-H. Additionally, we provide CD\text{-}$\ell_{2}$ results for the five categories with the largest training samples: table, chair, airplane, car, and sofa, each with over 2,500 samples. In contrast, we also present the CD$\text{-}\ell_{2}$ results for the five categories with the fewest training samples (birdhouse, bag, remote, keyboard, and rocket), each containing fewer than 80 samples. Our method consistently outperforms others across various conditions and perspectives. These findings highlight the model's exceptional ability to capture 3D shape information.

Figure~\ref{fig:9} illustrates the point cloud completion results from easy to hard on the ShapeNet-55 dataset. In the easy mode, our method can complete the missing point cloud without altering the overall shape. Even with significant missing information, our method still reconstructs a complete shape in the moderate and challenging modes, whereas PoinTr~\cite{yu2021pointr} fails to achieve satisfactory results. This demonstrates the effectiveness of our approach.


\begin{figure*}
	\centering
	\includegraphics[width=1.5\columnwidth]{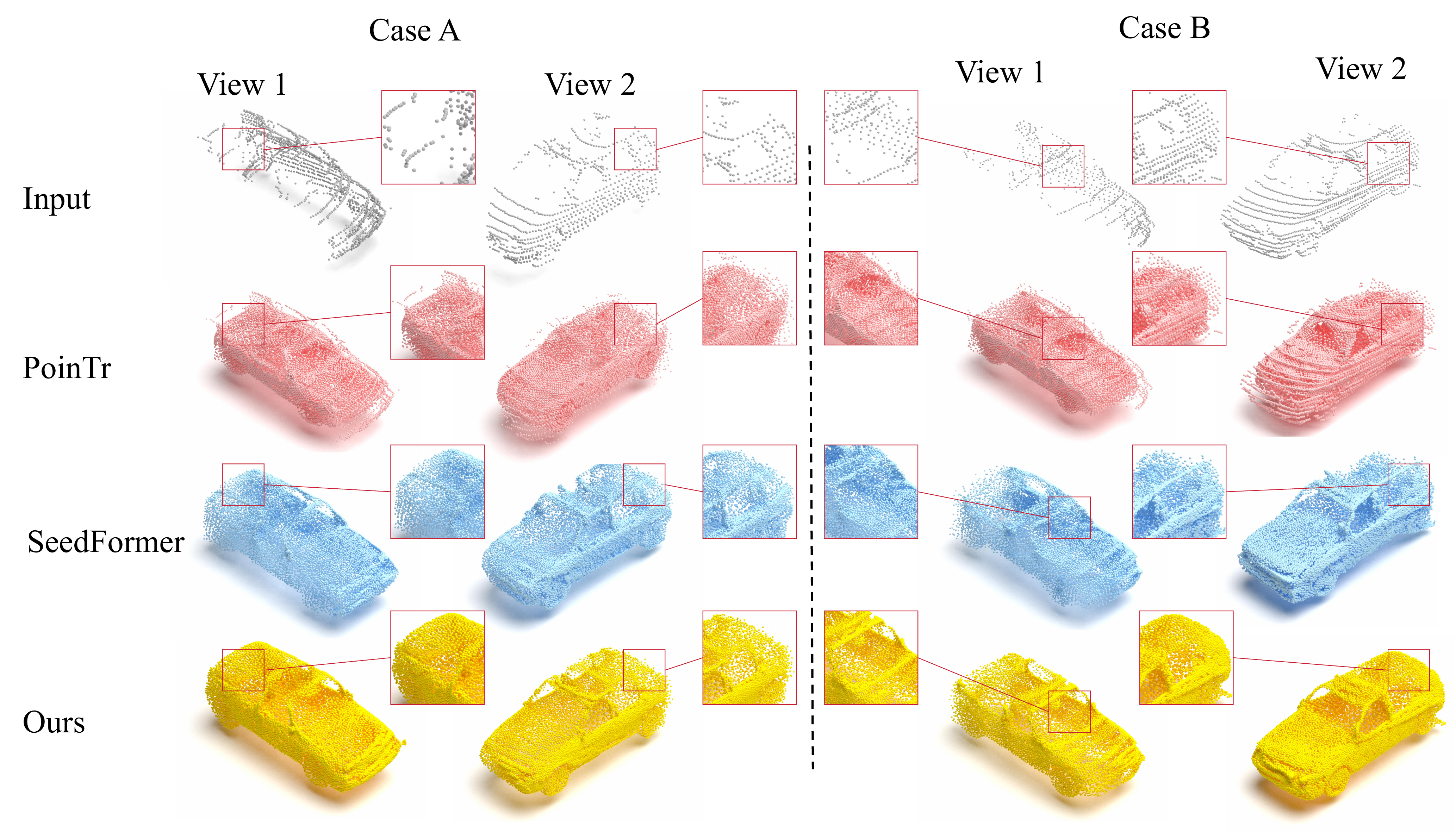}
	\caption{Qualitative results on KITTI dataset. We propose point cloud completion results of two car shapes in two different views.}
	\label{fig:12}
\end{figure*}

\begin{table}
	\centering  
	\begin{threeparttable}
		\caption{Quantative comparison on KITTI dataset in terms of FD and MMD.}
		\begin{tabularx}{\columnwidth}{c|>{\centering\arraybackslash}c>{\centering\arraybackslash}X>{\centering\arraybackslash}c>{\centering\arraybackslash}c|c}
			\toprule
			& PCN & SnowFlake & SeedFormer & PointAttN & Ours \\
			\midrule
			FD$\downarrow$  &2.235 & 0.110 & 0.151 & 0.672 & \textbf{0.097}\\
			MMD$\downarrow$ &1.366 & 0.907 & 0.516 & 0.504 & \textbf{0.463} \\
			\bottomrule
		\end{tabularx}
		\label{tb:4}
	\end{threeparttable}
\end{table}

\textbf{Evaluation on ShapeNet-34.} The ShapeNet-34 benchmark~\cite{yu2021pointr} rigorously tests the generalization capability of models by evaluating their performance on 21 previously unseen categories. This ensures that the models are not merely overfitting to known categories but can effectively generalize to novel, unseen objects, providing a comprehensive assessment of their robustness and adaptability in handling diverse 3D shapes. Table \ref{tb:3} presents a comprehensive analysis of the CD$\text{-}\ell_{2}$ performance for both seen categories and unseen classes. This is notable given the substantial shape variations between the training dataset and the unseen test data. Our DFG-PCN not only achieves the best average CD$\text{-}\ell_{2}$ performance across both types of objects but also demonstrates a generalization ability that surpasses other methods~\cite{zhou2022seedformer,xiang2021snowflakenet} by nearly 8.9\%. Additionally, DFG-PCN significantly enhances performance at the CD-H level, even when some inputs contain only 25\% of unseen objects.

As shown in Figures~\ref{fig:10} and~\ref{fig:11}, we present the visualization results for both seen and unseen categories. Our completion results remain outstanding, especially when the missing shapes are regular. Our results are very close to the ground truth. This indicates that the method demonstrates strong performance and generalization ability.

\begin{figure}
	\centering
	\includegraphics[width=1\columnwidth]{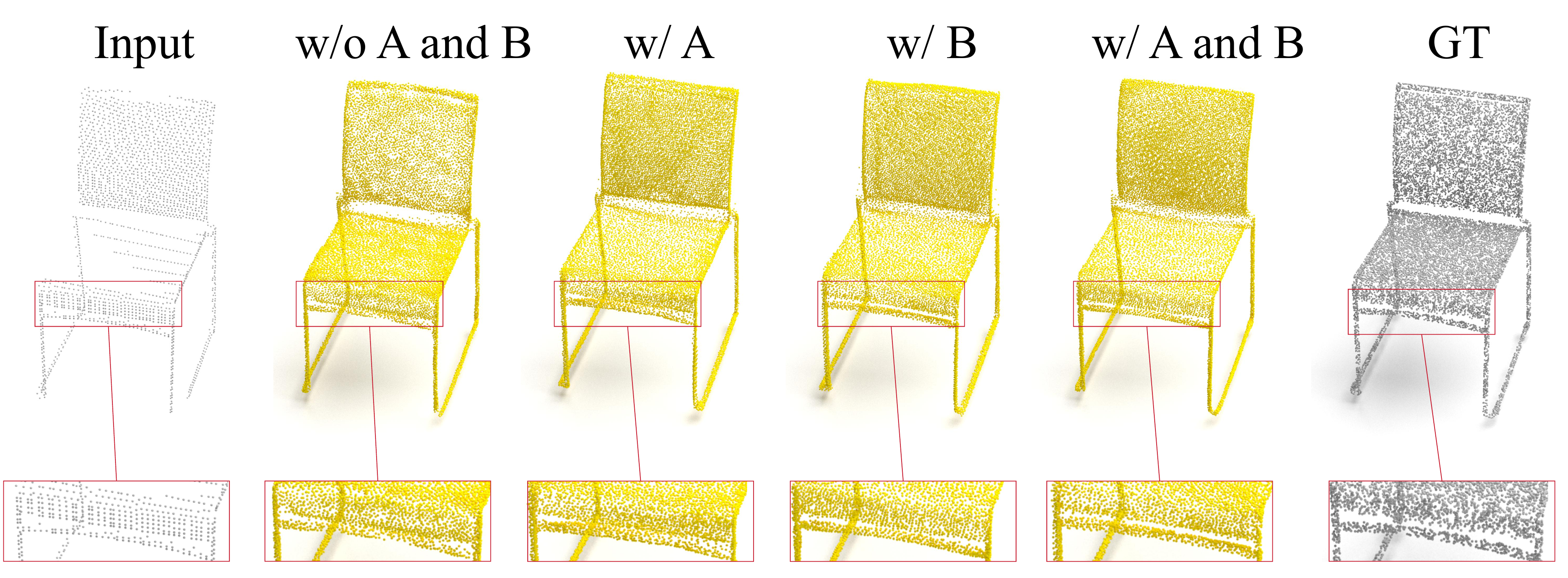}
	\caption{Visual comparison of different model variants. Adding geometric features (A) and the graph fusion module (B) leads to more accurate and detailed reconstructions.}
	\label{fig:13}
\end{figure}

\textbf{Evaluation on KITTI.} We further evaluate our approach on real-world datasets. As shown in Table \ref{tb:4}, our method consistently achieves superior performance compared to other methods in terms of both FD and MMD, demonstrating its strong ability to capture the 3D geometric characteristics of vehicles.

Additionally, qualitative results in Figure \ref{fig:12} present two completion examples from different viewpoints. Our method produces more accurate and visually coherent completions with finer local details and clearer structural patterns. 

\begin{table}
	\centering  
	\begin{threeparttable}
		\caption{Ablation study between KNN and Degree-Flexible Point Graph on the PCN dataset.}
		\begin{tabularx}{\columnwidth}{*2{>{\centering\arraybackslash}X}}
			\toprule
			Variations & CD\text{-}$\ell_{1}$ \\
			\midrule
			KNN  &6.56 \\
			Degree-Flexible & \textbf{6.42} \\
			\bottomrule
		\end{tabularx}
		\label{tb:5}
	\end{threeparttable}
\end{table}

\subsection{Ablation study}


\begin{table}
	\centering  
	\begin{threeparttable}
		\caption{Ablation study of local and global graph usages on the PCN dataset.}
		\begin{tabularx}{\columnwidth}{*4{>{\centering\arraybackslash}X}}
			\toprule
			 local graph & global graph &CD\text{-}$\ell_{1}$ \\
			\midrule
			\checkmark & &6.53 \\
			 & \checkmark & 6.49 \\
			 \checkmark & \checkmark &\textbf{6.42}     \\
			  	\bottomrule
		\end{tabularx}
		\label{tb:6}
	\end{threeparttable}
\end{table}

In this section, we present a series of ablation studies designed to evaluate the effectiveness of the proposed operations. All experiments are conducted under consistent settings on the PCN dataset~\cite{yuan2018pcn} to ensure fair comparisons. By systematically removing or modifying specific components of our model, we aim to assess their individual contributions to the overall performance, providing deeper insights into the impact of each operation on the shape completion task.

\textbf{Effectiveness of Degree-Flexible Point Graph.}
In typical applications of point cloud tasks, KNN graphs are commonly used for graph construction. However, KNN has its limitations, as all nodes are connected to a fixed number of neighboring nodes. To enhance flexibility, we propose an adaptive graph construction scheme that accounts for the unbalanced point demands in point cloud completion tasks. This scheme dynamically adjusts the connection structure based on the varying needs of different nodes, effectively catering to the diverse information requirements of each node in the graph. We compare the performance of KNN and DFG under the same settings. Table \ref{tb:5} demonstrates the improvements. Notably, our method exhibits significant and direct improvements over the KNN approach, where the CD\text{-}$\ell_{1}$ is reduced from 6.42 to 6.56, indicating a relative improvement of 2.1\%. This ablation study demonstrates that our method generates point clouds that are capable of retaining finer and more detailed shape information.

\textbf{Effectiveness of Global and Local Graphs.} In the design of DFG, we aggregate information from both local and global scales through local and global point node sampling. As shown in Table \ref{tb:6}, we conduct experiments using only local or global scale graphs under the same settings and compare them with the experiments of the original version of DFG. We remove the global graph or the local graph separately. However, both perform worse compared to global graph and local graph, highlighting the importance of the combination of local and global graph aggregation.

\textbf{Effectiveness of Detail Richness Metric.} To validate the effectiveness of our detail richness metric, we compare it with a classical geometric descriptor—curvature. Although curvature reflects local surface variation and is widely used in 3D geometry tasks, it lacks awareness of feature-space complexity and semantic importance. As shown in Table \ref{tb:7}, when using curvature alone to guide the graph degree allocation, the model achieves a CD\text{-}$\ell_{1}$ of 6.67 on the PCN dataset. In contrast, using our detail richness metric results in a lower CD\text{-}$\ell_{1}$ of 6.48, demonstrating a clear improvement in reconstruction quality. This performance gap suggests that the detail richness metric offers superior guidance, especially in identifying semantically complex or structurally incomplete regions that may not exhibit strong geometric curvature. Unlike curvature, which only responds to geometric shape variation, our feature-based metric captures learned feature inconsistency and geometric information, providing a more comprehensive signal for adaptive graph construction.

\textbf{Effectiveness of Geometric Feature and Graph Fusion.}
We conducted an ablation study to validate the effectiveness of the geometric feature and Graph Fusion Module (GFM). As shown in Table \ref{tb:8}, removing both geometric features and GFM yields a baseline CD\text{-}$\ell_{1}$ of 6.48. Individually incorporating geometric features or GFM reduces the CD\text{-}$\ell_{1}$ to 6.46 and 6.44, respectively, highlighting their contributions in enhancing local detail perception and integrating multi-scale features. Utilizing both modules simultaneously achieves the best performance. Additionally, the visualization results in Figure \ref{fig:13} clearly illustrate that models equipped with geometric features and GFM reconstruct more accurate and detailed local structures, demonstrating their complementary roles in improving completion quality and structural consistency.

\begin{table}
	\centering  
	\begin{threeparttable}
		\caption{Comparison of curvature and detail richness metric on the PCN dataset.}
		\begin{tabularx}{\columnwidth}{*2{>{\centering\arraybackslash}X}}
			\toprule
			Metrics & CD\text{-}$\ell_{1}$ \\
			\midrule
			Curvature  &6.67 \\
			Detail Richness Metric & \textbf{6.48} \\
			\bottomrule
		\end{tabularx}
		\label{tb:7}
	\end{threeparttable}
\end{table}

\begin{table}
	\centering  
	\begin{threeparttable}
		\caption{Ablation study of geometric feature and graph fusion usages on the PCN dataset.}
		\begin{tabularx}{\columnwidth}{*3{>{\centering\arraybackslash}X}}
			\toprule
			geometric feature  &graph fusion &CD\text{-}$\ell_{1}$ \\
			\midrule
			& &6.48  \\
			\checkmark & &6.46 \\
			& \checkmark & 6.44 \\
			\checkmark & \checkmark &\textbf{6.42}     \\
			\bottomrule
		\end{tabularx}
		\label{tb:8}
	\end{threeparttable}
\end{table}

\section{Limitation and Future Work}\label{lf}
Our method has a few limitations that open up possibilities for future improvements. First, while a flexible graph structure for point cloud processing has been implemented, there is still room for optimization in how the graph is constructed. Enhancing the graph's ability to establish effective connections and ensure smooth information transfer between nodes will be a priority. Future research will focus on devising adaptive algorithms for graph connections that can adjust according to the distinct shape characteristics of point clouds. Secondly, since we currently handle each dataset independently for training and prediction, the ability to generalize across datasets remains an issue. Some datasets' small sizes and differences limit the model's ability to be widely applicable. To address this, future efforts will be directed at developing methods to improve cross-dataset generalization and strengthen the robustness of the model overall.
\section{Conclusion}\label{cl}
In this paper, we present a novel point cloud completion method designed to alleviate the bottleneck of two-stage frameworks, with a particular focus on the second stage. Our approach introduces a newly designed flexibility graph to recover fine-grained shape information for the missing parts. It enhances the ability to represent seed points by aggregating global and local graphs. The experimental results highlight the benefits of our method, while qualitative evaluations of point cloud completion showcase its capability to represent intricate geometric details and effectively reconstruct missing areas accurately. Expanding this approach to related tasks, such as point cloud reconstruction and upsampling, offers a promising direction for future research.

\section*{Acknowledgments}\label{sec:acknowledgements}
This work is supported by the National Natural Science Foundation of China (62172356, 61872321, 62272277), Zhejiang Provincial Natural Science Foundation of China (LZ25F020012),
the Ningbo Major Special Projects of the ``Science and Technology Innovation 2025'' (2020Z005, 2020Z007, 2021Z012).

\ifCLASSOPTIONcaptionsoff
  \newpage
\fi

\bibliographystyle{IEEEtran}
\bibliography{reference}



\begin{IEEEbiography}[{\includegraphics[width=1in,height=1.25in,clip,keepaspectratio]{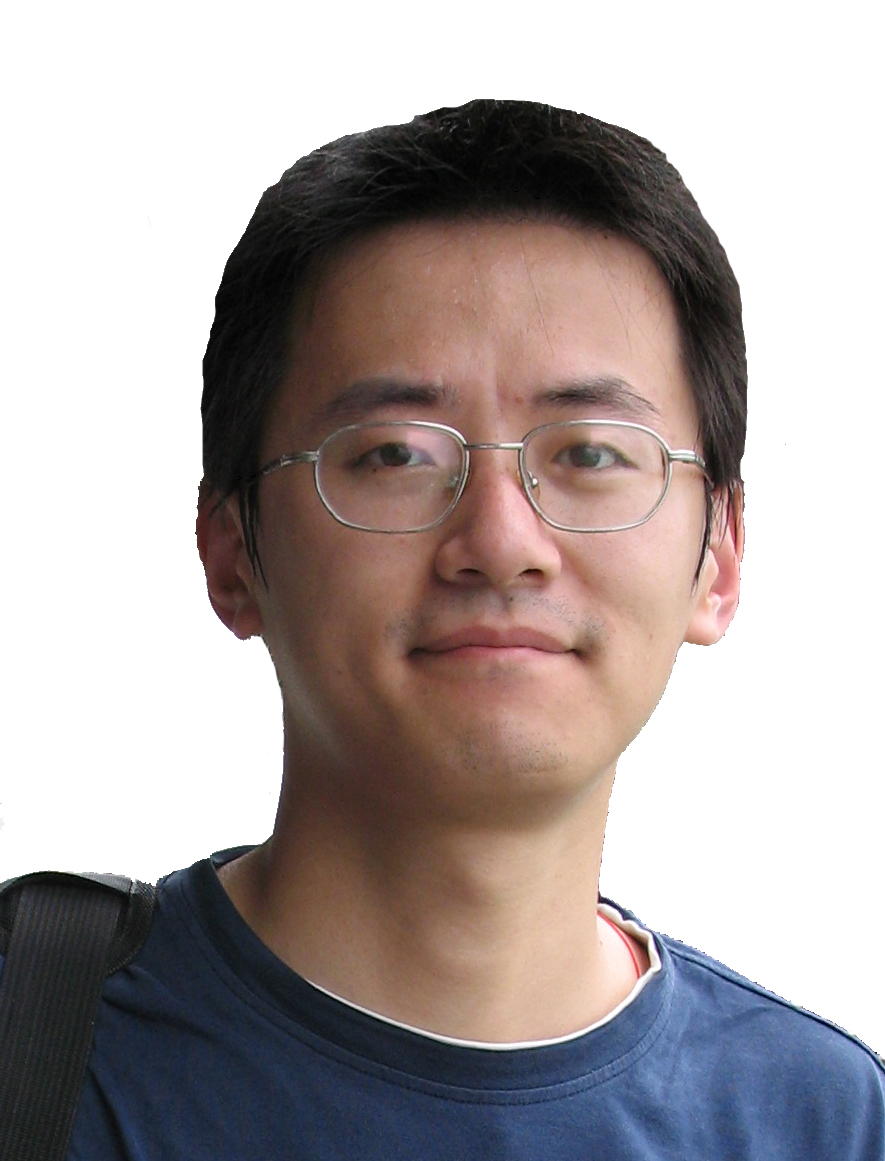}}]{Zhenyu Shu}
got his Ph.D. degree in 2010 at Zhejiang University, China. He is now working as a full professor at NingboTech University. His research interests include computer graphics, digital geometry processing and machine learning. He has published over 30 papers in international conferences or journals.
\end{IEEEbiography}

\begin{IEEEbiography}[{\includegraphics[width=1in,height=1.25in,clip,keepaspectratio]{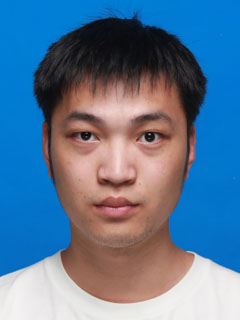}}]{Jian Yao}
	is a graduate student of the School of  Computer Science and Technology at Zhejiang Sci-Tech University. His research interests include computer graphics and machine learning.
\end{IEEEbiography}

\begin{IEEEbiography}[{\includegraphics[width=1in,height=1.25in,clip,keepaspectratio]{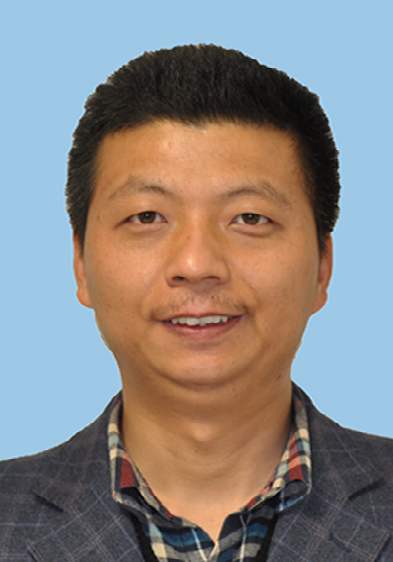}}]{Shiqing Xin}
	is an associate professor at the Faculty of School of Computer Science and Technology in Shandong University. He received his Ph.D. degree in applied mathematics at Zhejiang University in 2009. His research interests include computer graphics, computational geometry and 3D printing.
\end{IEEEbiography}


\vfill







\end{document}